\documentclass[11pt, final]{article}
\usepackage{MRnotes}

\newcommand{\slt}{\mathfrak{sl}(2)}
\newcommand{\sut}{\mathfrak{su}(2)}
\newcommand{\psut}{\mathfrak{psu}(1,1|2)}
\newcommand{\Dta}{\mathrm{D}(2,1|\alpha)}

\newcommand{\Gt}{\widetilde{G}}

\newcommand{\asx}[1]{\mathrm{AdS}_3 \times \mathbf{S}^3 \times #1}
\newcommand{\asth}{\mathrm{AdS}_3 \times \mathbf{S}^3}
\newcommand{\stso}{\mathbf{S}^3 \times \mathbf{S}^1}
\newcommand{\scg}{\bm{\chi}_{_\text{g}}}
\newcommand{\scgb}{\overline{\bm{\chi}}_{_\text{g}}}
\newcommand{\schr}[1]{\bm{\chi}_{#1}}

\newcommand{\sccft}{\bm{\xi}}

\title{One-loop determinants in \AdS{3} supergravity with extended supersymmetry}
\author{Ilija Rakic, Lorenzo Toni}

\affiliation{
	Center for Quantum Mathematics and Physics (QMAP)\\
	Department of Physics \& Astronomy, University of California, Davis, CA 95616 USA}

\emailAdd{irakic@ucdavis.edu}
\emailAdd{ltoni@ucdavis.edu}

\abstract{We re-examine the computation of the one-loop partition function of Type II supergravity theory compactified on $AdS_3 \times \mathbf{S}^3 \times X$, where $X$ can be $K_3$, $T^4$ and $\mathbf{S}^3 \times \mathbf{S}^1$. These backgrounds preserve eight supercharges (four left and four right moving) and are known to be holographically dual to either small or large $\mathcal{N}=(4,4)$ superconformal field theories. By extending well-established heat kernel techniques, we evaluate the one-loop determinants associated to these supergravity backgrounds. The resulting expressions are then used to reconstruct the characters of short multiplets of the corresponding dual boundary theories. A distinctive aspect of our analysis is the emergence of contributions from multiple gravitational saddle points in the path integral, reflecting the spectral flow structure present in the boundary theory.}

\begin{document} 
\maketitle


\section{Introduction}\label{sec:intro}

Three-dimensional gravity with a negative cosmological constant serves as one of the most fruitful models in the study of quantum gravity. A wide literature on the gravitational physics in this spacetime is available and encompasses microscopic descriptions, such as the holographic $\text{AdS}\text{/}\text{CFT}$ correspondence and the exactly solvable models of String Theory, but also the physics of black holes, which in three dimensions only exist in negatively curved spacetime. The subject of this paper will be semiclassical gravity on \AdS{3}, which has a simplifying feature unique to three dimensions, and that is the fact that there are no graviton excitations in the bulk and the perturbative dynamics is captured by non-trivial boundary degrees of freedom.

One way to infer the existence of these boundary degrees of freedom is through the analysis of~\cite{Brown:1986nw}, who showed that these modes are associated with the asymptotic symmetry algebra of the theory, which itself is the algebra that the isometry algebra of the bulk geometry gets enlarged to on the boundary. The simplest example is provided by pure gravity, i.e. gravity without additional fields, where the isometry group $\mathrm{SL}(2, \mathbb{R}) \times \mathrm{SL}(2, \mathbb{R})$ gets enlarged to two copies of the centrally extended Virasoro algebra with central charge $c = \frac{3\,\lads}{2\,G_N}$. Based on this observation, it was further argued in~\cite{Maloney:2007ud} that the graviton partition function, or equivalently, its one-loop determinant around the \AdS{3} geometry, should reproduce the vacuum character of the two Virasoro algebras. This result was then directly derived in semiclassical gravity using heat-kernel techniques in~\cite{Giombi:2008vd}.

If instead supergravity theories in \AdS{3} are considered, the isometry algebra gets extra fermionic generators, which promotes it to a superisometry algebra, and its associated asymptotic algebra is promoted to a superconformal algebra~\cite{Henneaux:1999ib}. Physically, this amounts to a richer spectrum of boundary modes. The analysis of~\cite{Maloney:2007ud} now applies to the partition function of the full spectrum of the supergravity theory, and the argument is that it should reproduce the vacuum character of the superconformal algebra. This has been verified in the case of $\mathcal{N}=1$ supergravity in~\cite{David:2009xg} using the aforementioned heat-kernel techniques to find the one-loop determinants of the gravitons and gravitini of this theory. For theories of extended supergravity, the same procedure is expected to work.

In this paper, we will show that this is indeed the case. The same strategy will be employed, but there is an additional important subtlety that needs to be addressed. Since the modules of extended superconformal algebras have non-trivial null states, their characters are more involved. It turns out that, once all the determinants of all the fields in the supergravity multiplet are accounted for, only a piece of this character is obtained. We proceed by realizing that the superconformal character can be obtained from this seed piece by implementing an algebra operation known as spectral flow. It is then natural to ask what this operation corresponds to in the gravitational analysis. As described in~\cite{Henneaux:1999ib,Dijkgraaf:2000}, and carefully analyzed in~\cite{Kraus:2006nb}, the theories with extended supergravity involve background R-charge gauge fields. The boundary conditions for these gauge fields single out a whole set of configurations, and spectral flow amounts to switching between these configurations. These gauge fields have Chern-Simons dynamics, and do not backreact on the geometry, but since the gravitini and other fields in the spectrum carry R-charge, their one-loop determinant will be different in each of these backgrounds, with the seed element being one of those backgrounds. The path integral sum over geometries with different gauge field configurations leads, at the end of the day, to an answer that agrees with the superconformal character. This calculation is summarized by the following equation involving the partition function:
\begin{equation}\label{eq:calcsumm}
    Z_{\text{SUGRA}} = \sum_{m \, \in \, \text{\{Configurations\}}} \, \, \prod_{i \, \in \, \text{\{Spectrum\}}} Z_{\text{i-th field in m-th conf.}} \, .
\end{equation}
This idea is well-appreciated by experts. For instance, the recent calculation of one-loop determinants in super-JT gravity to analyze the thermodynamics of near-BPS black holes~\cite{Heydeman:2020hhw,Heydeman:2025a,Heydeman:2025b} demonstrates this explicitly. Likewise, the relevant characters were recently derived using string worldsheet techniques~\cite{Ferko:2024uxi,Murthy:2025}. We will simply perform the calculation directly in $AdS_3$.

We will perform our analysis for geometries of the form $\asx{X}$.\footnote{
    For definiteness, we focus on Type IIB superstrings. The geometries in question arise as the near-horizon spacetime of five-brane and one-brane bound states.
} For $X=K3$ or $T^4$, after dimensional reduction on the compact spaces, the three-dimensional supergravity has a superisometry algebra $\psut \times \overline{\psut}$, which gets enlarged to the \emph{small} $\mathcal{N} = (4,4)$ superconformal algebra on the boundary.\footnote{
    The distinction between $K3$ and $T^4$ is that the latter additionally has 4 left moving and 4 right moving $\mathrm{U}(1)$ currents, and their fermionic partners, while preserving the same amount of supersymmetry.}
On the other hand, $X = \stso$ leads to the superisometry algebra $\Dta \times \overline{\Dta}$, which turns into the \emph{large} $\mathcal{N} =(4,4)$ superconformal algebra. We will use the supergravity spectrum obtained in~\cite{Deger:1998nm},~\cite{Larsen:1998xm}, and~\cite{Eberhardt:2017fsi}, respectively, for the three cases. Furthermore, we will ignore the massive states arising from the Kaluza-Klein reduction and focus directly on the light states in pure supergravity. The relevant supergravity theories were constructed in~\cite{Henneaux:1999ib}.

We will work in the Euclidean \AdS{3} geometry. Furthermore, we will pick twisted boundary conditions around the $\mathbf{S}^3$(s) to turn on $\mathrm{SU}(2)$ R-charge chemical potentials. Since our theories have fermionic fields, we need to specify a spin structure for computing one-loop determinants. Without loss of generality, we will take periodic boundary conditions for fermions around the Euclidean time circle, which in the boundary superconformal algebra correspond to the signed Neveu-Schwarz sector trace with the $(-1)^F$ insertion. The trace without the insertion, which corresponds to antiperiodic fermion boundary conditions, can then be obtained by tuning the R-charge chemical potential to suitable values.

The plan of the paper is the following. We start~\cref{sec:K3onel} by introducing the isometry algebra of $\asth$, the $\psut \times \overline{\psut}$, and its associated asymptotic algebra, the $\emph{small}$ $\mathcal{N} = (4,4)$ superconformal algebra. In~\cref{sec:K3spec} we present the spectrum of 3D gravity obtained from dimensional reduction on the internal space of $\asx{K3}$, and we write this in terms of the characters of the isometry group. We then turn to heat kernel techniques in~\cref{sec:hkgeneral}, which we use to compute the one-loop determinant of a field fluctuating in $\asth$, which is equivalent to the one-loop determinant in \AdS{3} with background gauge fields turned on. From the one-loop determinant, we then find the partition function. This generalizes known results to account for these background Chern-Simons gauge fields. We also describe the "multi-particling" technique, which is a quick alternative method of finding the partition function of a field, directly from the isometry algebra characters associated with that field. In~\cref{sec:sf} we show how boundary conditions for the background gauge fields generate a whole set of gauge configurations, all of which need to be taken into account when computing the full partition function. Furthermore, we define the spectral flow of the dual conformal field theory, and show that this operation has an interpretation in the bulk in terms of these gauge fields. In~\cref{sec:K3char} we put together the partition functions of all of the massless fields in the spectrum, and perform the path integral over the gauge field configurations, which gives us the partition function of \AdS{3} supergravity obtained from dimensional reduction on $\asx{K3}$. We compare this to the product of the superconformal character of the dual boundary algebra, and show that they agree. We end~\cref{sec:K3onel} by repeating the same analysis for $\asx{T^4}$, which includes additional fields in the spectrum. In~\cref{sec:S3S1onel} we deal with the $\asx{\mathbf{S}^3 \times \mathbf{S}^1}$ background, which has a more intricate symmetry algebra structure determined by the relative sizes of the two $\mathbf{S}^3$ factors. We begin with the isometry group $\Dta \times \overline{\Dta}$ and associated asymptotic algebra, the $\emph{large}$ $\mathcal{N} = (4,4)$ superconformal algebra. Then we introduce the spectrum in~\cref{sec:S3S1spec}, generalize the heat kernel and spectral flow methods in~\cref{sec:S3S1hk}, and put together the full partition function in~\cref{sec:S3S1char}, which we again compare to the superconformal characters. We conclude with a discussion of how our method for calculating the partition function fits into and contrasts recent calculations of the same object, using different methods, and how the thermal \AdS{3} partition function ties to black hole physics.~\cref{sec:algebras} discusses the two $\mathcal{N}=4$ superconformal algebras we use, and~\cref{sec:hkapp} compiles all of the partition functions we use throughout this work.

\section{One-loop supergravity determinant around \texorpdfstring{$\asx{K3}$}{AdS3K3}}\label{sec:K3onel}

Consider for definiteness Type IIB supergravity compactified down to six-dimensions on a Calabi-Yau two-fold $K3$, preserving 6D $\mathcal{N}=(2,0)$ supersymmetry (eight supercharges). These 6D theories have $\asth$ vacua, supported by either NS-NS or R-R two-form flux, or a combination thereof, still preserving the same eight supercharges.

The superisometry algebra of $\asth$ consists of two copies of $\psut$~\cite{deBoer:1998kjm}, each of which is generated by $\{L_0,L_{\pm1},J_0^i,G_{\pm\frac{1}{2}}^a,G_{\pm\frac{1}{2}}^{b}\}$, where $i=1,2,3$ and $a,b=1,2$ are $\sut$ vector and spinor indices, respectively. The commutation relations can be found in~\cref{sec:smallalg}. We will focus on one copy of $\psut$. It has bosonic subalgebras $\slt$ and $\sut$, which are generated by $\{L_0,L_\pm\}$ and $J_0^i$, respectively. These two, when combined with the other copy, correspond to the isometries of the \AdS{3} and $\mathbf{S}^3$ factors. The remaining generators, $\{G_{\pm\frac{1}{2}}^a,G_{\pm\frac{1}{2}}^{b}\}$, are the supercharge fermionic operators, which transform as an $\sut$ doublet and its conjugate.

A representation of this algebra is labeled by the weight and spin $(h,j)$ of the highest weight state $\ket{h,j}$ defined as one annihilated by the lowering operators $ \{L_{+1},J_0^+,G_{+\frac{1}{2}}^a, G_{+\frac{1}{2}}^{b}\}$. From this state, we obtain the rest of the representation by acting with the raising operators $\{L_{-1},J_0^-,G_{-\frac{1}{2}}^a, G_{-\frac{1}{2}}^{b}\}$. Unitary representations satisfy the bound
\begin{equation}\label{eq:smallunitbnd}
    h\geq j \, ,
\end{equation}
with the inequality being saturated by the short representations. Since generically supergravity only sees chiral primary BPS states, which saturate the unitarity bound and lead to short multiplets, we will exclusively focus on them. The representation content of the algebra can be succinctly encapsulated into a character, which will be given in~\cref{sec:K3spec}.

Due to the asymptotics of \AdS{3}, $\psut\times\overline{\psut}$ gets enhanced to a pair of $\mathcal{N} = 4$ superconformal asymptotic algebras~\cite{deBoer:1998kjm} with central charges $c = \frac{3\,\lads}{2\,G_N}$. Focusing on the holomorphic part, the generators are $\{L_n, J^i_n, G_r^a, G_r^{b}\}$ where $n \in \mathbb{Z}$, $i \in \{1,2,3\}$ and $a,b\in \{1,2 \}$. Furthermore, we will be interested in $r \in \mathbb{Z}+\frac{1}{2}$, the Neveu-Schwarz sector ($r\in \mathbb{Z}$ is the Ramond sector). The commutation relations are given in~\cref{sec:smallalg}.

The representation theory of this algebra closely follows that of the isometry algebra. Representations are labeled by two quantum numbers $(h,\ell)$ associated with $L_0$ and $J_0^3$.\footnote{
We will be using different spin labels for the asymptotic and isometry algebras, $\ell\neq j$, as they will not be equal in our analysis. Once we start matching the partition function to the character, the supergravity spectrum, associated with $j$, will instruct us what values of $\ell$ to take for the character.
} A highest weight state $\ket{h,\ell}$ is as usual annihilated by the positively moded generators and lies in the spin-$\ell$ highest weight representation of $\sut$.
The rest of the representation can be obtained by acting with the negatively moded generators. This produces a collection of states among which certain linear combinations, the null vectors, need to be removed (see~\cref{sec:smallalg}). Unitary representations in the NS-sector are constrained by having the $\sut$ spin satisfy the same bound $h \geq \ell$. Representations saturating this bound are the short BPS multiplets. The superconformal characters of the asymptotic algebra have been compiled in~\cref{sec:smallalg}, and we will quote them from there once they are needed.

\subsection{$\psut$ and the spectrum in three dimensions}\label{sec:K3spec}

The 6D $\mathcal{N} = (2,0)$ SUGRA theory obtained by compactifying IIB SUGRA theory on $K3$ has a graviton multiplet and $21$ tensor multiplets, with a global $\mathrm{SO}(5) \times \mathrm{SO}(21)$ symmetry. Decomposing these fields into $\mathbf{S}^3$ harmonics, the resulting three-dimensional spectrum can be assembled in terms of representations of $\psut \times \overline{\psut}$, as worked out in detail in~\cite{Deger:1998nm,deBoer:1998kjm}.

We will summarize the spectrum in terms of the characters of this algebra, which are defined for one copy of $\psut$ as the following trace over the representation space\footnote{We will sometimes refer to these as the gravitational characters, and label them with $\schr{}^\text{g}$, as they are associated with the supergravity spectrum, while the characters of the superconformal algebra will be labeled with $\schr{}$.}
\begin{equation}\label{eq:psu2chardef}
    \schr{}^\text{g}(h,j) =
    \text{Tr}{_\text{}}\left((-1)^{F} \, 
    q^{L_0} \, z^{J_0^{3}} \right)\,,
\end{equation}
where $q = e^{2\pi i\, \tau}$ and $z= e^{2\pi i \nu}$ are related to the asymptotic boundary modular and elliptic parameter, and will be discussed in~\cref{sec:hkgeneral}. The fermionic number operator $F$ is inserted to account for periodic boundary conditions for the fermions. For a short representation, the character evaluates to
\begin{equation}\label{eq:psu2char}
\schr{}^\text{g}(j) = q^j\, \chi_j(z) - 2\, q^{j+\frac{1}{2}}\, \chi_{j-\frac{1}{2}}(z) + q^{j+1}\, \chi_{j-1}(z)\,,
\end{equation}
where $\chi_j(z)$ is the $\sut$ Lie algebra character
\begin{equation}\label{eq:su2char}
        \chi_j(z) = 
        \dfrac{z^{j+1/2} - z^{-j-1/2}}{z^{1/2} - z^{-1/2}}\,,
        \qquad
        j \geq 0 \, .
\end{equation}
For small values of $j$ there are truncations in the representations which are captured by the following characters
\begin{equation}\label{eq:psu2chartrunc}
    \begin{split}
    \schr{}^\text{g}(0) & = q^0 \chi_0(z) = 1 \, ,
    \qquad
    \schr{}^\text{g}(1/2)= q^{1/2} \chi_{1/2}(z) + 2 \, q^1 \chi_0(z) \, , \\
     & \schr{}^\text{g}(1) = q^1\chi_1(z) - 2q^{3/2} \chi_{1/2}(z) + q^2\chi_0(z) \, .
    \end{split}
\end{equation}

To give a compact presentation of the supergravity spectrum, we combine left and right moving $\psut$ characters into combinations specified by definite helicity $s=h-\overline{h}$ of the highest weight state, and we denote them as $\sccft^{(s)}(j)$. These global characters are defined as follows:
\begin{equation}\label{eq:helchar}
\begin{aligned}
\sccft^{(2)}(j) 
&=   
	\schr{}^\text{g}(j+1)\; \overline{\schr{}}^\text{g}(j) \,, 
&  \quad
\overline{\sccft}^{(2)}(j)
&=   
	\schr{}^\text{g}(j)\; \overline{\schr{}}^\text{g}(j+1)\, ,   \\ 
\sccft^{(\frac{3}{2})}(j) 
&= 
	\schr{}^\text{g}(j+\frac{1}{2})\; \overline{\schr{}}^\text{g}(j) \,, 
&  \quad 
\overline{\sccft}^{(\frac{3}{2})}(j) 
&= 
	\schr{}^\text{g}(j)\; \overline{\schr{}}^\text{g}(j+\frac{1}{2})\,,\\
& \hspace{2.6cm} 
    \sccft^{(1)}(j) 
=&  
	\hspace{-2mm} \schr{}^\text{g}(j)\; \overline{\schr{}}^\text{g}(j) \,.   
\end{aligned}
\end{equation}
One can check using~\eqref{eq:psu2char} that, for example, in $\sccft^{(2)}(j)$ the highest weight $h$ is $j+2$ and lowest $\bar{h}$ is $j$, giving $s=2$. The $\bar{\sccft}$ on the right are complex conjugates of $\sccft$, which give representations with the opposite helicity.

The analysis of~\cite{Deger:1998nm,deBoer:1998kjm} gives the spectrum of the 6D supergravity theory to consist of the following multiplets:
\begin{itemize}[wide,left=0pt]
\item  A spin-$2$ multiplet with character $\sccft^{(2)}(j)$ for $j \geq \frac{1}{2}$. Note that the spin-$2$ multiplet $\sccft^{(2)}(0)$ corresponding to the boundary supergravitons is not visible in the analysis of~\cite{Deger:1998nm}, owing to their gauge fixing conditions. We will therefore extend the range to $j\geq 0$.   
\item A single spin-$1$ multiplet from 6D supergraviton multiplet $\sccft^{(1)}(j)$ for $j\geq 1$. 
\item 21\footnote{
    As far as supersymmetry is concerned, we can have an arbitrary number $n$ of these spin-$1$ vector multiplets, but $n=21$ is required for anomaly cancelation.
} spin-$1$ multiplets living in the vector of $\mathrm{SO}(21)$ $\sccft^{(1)}(j)$ for $j\geq 1/2$ obtained from the 6D tensor multiplets. 
\end{itemize}
Using ($\ref{eq:helchar}$), we can encode the spectrum using the character decomposition as
\begin{equation}\label{eq:sugraSMK3}
\begin{split}
& 
    \sccft^{(2)}(0) 
    + \sum_{j \geq \frac{1}{2}} \left[
    \sccft^{(2)}(j) +  
    \mathbf{21}_{21}\, \sccft^{(1)}(j)  + 
    \sccft^{(1)}\left(j+\frac{1}{2}\right) 
    \right] + \text{cc.} \, ,
\end{split}
\end{equation}
where the $\text{cc.}$ refers to the representations with opposite helicity (given in terms of $\overline{\sccft}$).

We will be interested in the lowest level of this tower, $j=0$, which we have singled out in~\eqref{eq:sugraSMK3}. It is associated with the boundary graviton modes, but it will also include other boundary modes. The field content can be found by applying~\eqref{eq:psu2chartrunc} and~\eqref{eq:helchar}:
\begin{equation}\label{eq:gravchar}
    \sccft^{(2)}(0) = \left[ q^1\chi_1(z) - 2 \, q^{3/2} \chi_{1/2}(z) + q^2\chi_0(z) \right] \times \Big[(\bar{q}^0 \chi_0(\bar{z}))\Big].
\end{equation}
Each term in this equation has the form of $q^h \, \bar{q}^{\bar{h}} \, \chi_{\ell}\,(z) \, \chi_{\bar{\ell}}\,(\bar{z})$, from which the $(h,\bar{h},\ell,\bar{\ell})$ labels of each field can be extracted. For example, the first term represents a vector field ($s=h-\bar{h}=1$) which transforms as an $\sut$ triplet. Repeating this for each term we find that this multiplet consists of
\begin{equation}\label{eq:gravqnum}
    1 \times \Big( 1,0,1,0 \Big) \ + \ 2 \times \left( \frac{3}{2},0,\frac{1}{2},0 \right) \ + \ 1 \times \Big( 2,0,0,0 \Big),
\end{equation}
where the first term is the vector, the second two gravitini doublets, and the last term a graviton\footnote{
    As we mentioned earlier, this level is not accounted for by the tables provided by~\cite{Deger:1998nm}, but they can be seen by plugging in $\ell=-1$ in Table 3, and ignoring fields with unphysical values of $\sut$ spin. The quantum numbers of the remaining fields agree with what we have in~\eqref{eq:gravqnum}}.
In \AdS{3}, these fields correspond to asymptotic symmetries. The spin$-2$ field, or the boundary graviton modes, corresponds to large diffeomorphisms associated with Virasoro symmetry. The spin$-1$ holomorphic currents correspond to large gauge transformations of Chern-Simons fields associated with the isometries of the $\mathbf{S}^3$, which we will introduce later. The fermionic fields correspond to large SUSY transformations. Furthermore, one can check that each of these fields is massless using~\eqref{eq:mass_spin}.

\subsection{Heat kernel techniques in \AdS{3} with background gauge fields}
\label{sec:hkgeneral}

We now turn to the calculation of the one-loop determinants around $\asth$, which will then be used to calculate partition functions in this background. The calculations that were originally done separately for $\mathbf{S}^3$ and \AdS{3} in~\cite{David:2009xg} will serve as our building blocks. Once we obtain the result, we will argue that it can also be interpreted as a one-loop determinant around \AdS{3} with background Chern-Simons gauge fields that are obtained from compactification on the $\mathbf{S}^3$. Using this result, we will be able to write down the supergravity partition function for the cases with small $\mathcal{N} =4 $ superconformal symmetry, viz., for the $\asx{K3}$ and $\asx{T^4}$ compactifications.

We will work in Euclidean signature, and take the line-element of $\asth$ to be 
\begin{equation}\label{eq:6Dlelem}
ds^2_{_\text{6D}} = d\rho^2 + \cosh^2\rho\, d\tE^2 + \sinh^2\rho\, d\varphi^2 + d\theta^2 + \sin^2 \theta\, d\phi_1^2 + \cos^2\theta\, d\phi_2^2\,.
\end{equation}
We are setting the \AdS{3} scale to one. The Euclidean time coordinate, which we refer to as the thermal circle, is compact, and we implement the following identifications  $\varphi \sim \varphi +2 \pi$, $\phi_i \sim \phi_i + 2\pi$,  and impose twisted boundary condition
\begin{equation}\label{eq:6Dident}
(\tE, \varphi, \phi_1,\phi_2) \sim (\tE + \beta, \varphi +i\, \beta\, \mu, \phi_1 +i\, \beta\, \nu_1, \phi_2 +i\, \beta\,\nu_2)\,.
\end{equation}
Owing to these identifications, the $\mathbf{S}^3$ is non-trivially fibered over the \AdS{3}. The boundary of the spacetime is a $T^2$ with coordinates $(\tE,\varphi)$, whose complex structure is captured by the modular parameter
\begin{equation}\label{eq:tauads}
\tau = \frac{\beta\mu  + i\,\beta}{2\pi} \,, \qquad q\equiv e^{2\pi i\,\tau}\,.
\end{equation}
The twists along the Hopf angles of the $\mathbf{S}^3$ can be captured by the elliptic parameter
\begin{equation}\label{eq:nuads}
\nu = \beta\, \nu_2 - \beta\, \nu_1\,, \qquad z = e^{2\pi i\,\nu} \,,
\end{equation}
which as we will see can also be viewed as a complex chemical potential associated with the Chern-Simons gauge fields that appear when further reducing to \AdS{3}.

To compute one-loop determinants, we will use the heat kernel technique, which we now describe. The spin-$s$ heat kernel on $\asth$ is defined as
\begin{equation}\label{eq:hk}
K^{(s)}_{} (x,y;t) =  
\sum_{n}^{} \Psi_{n}^{(s)}(x) \Psi_{n}^{*(s)}(y) \ e^{t \lambda_n^{(s)}},
\end{equation}
where $\Psi_{n}^{(s)}(x)$ and $\lambda_n^{(s)}$ are respectively the eigenfunctions and eigenvalues of the spin-$s$ Laplace operator on this space, and $n$ is a composite index that runs over the entire spectrum of this operator. Since our space factorizes, the eigenfunctions are products of eigenfunctions on \AdS{3} and $\mathbf{S}^3$. Furthermore, since each factor has maximal symmetry, we can think of the index as running over representations of the underlying symmetry groups $\mathrm{SO}(4)$ and $\mathrm{SO}(2,2)$. The representations on $\mathbf{S}^3$ are labeled by two $\sut$ indices $(\ell,\bar{\ell})$, and the eigenfunctions, spherical harmonics, are known, so the heat kernel can be directly calculated. The relevant representations on \AdS{3} are also known~\cite{Camporesi:1995}, but the heat kernel on \AdS{3} can alternatively be obtained by analytically continuing the result for $\mathbf{S}^3$, which is how it was obtained in~\cite{David:2009xg}.

Since our space is a thermal quotient of $\asth$, meaning it is obtained from $\asth$ by imposing identifications~\eqref{eq:6Dident}, we can use the method of images to find the heat kernel on this quotient space as follows. Given a group manifold $G$ and a group of identifications $\Gamma$, the heat kernel on the quotient space $G/\Gamma$ is obtained by fixing one point $x$ in the original heat kernel and summing over all possible images $\gamma^{}(y)$ of the other point $y$::
\begin{equation}\label{eq:hkquot}
K_{G/\Gamma}(x,y;t) = \sum_{\gamma \in \Gamma} K^{}_{G} (x,\gamma^{}(y);t).
\end{equation}
The transformations $\gamma$ are coordinate shifts by the identification periods defined in~\eqref{eq:6Dident}, and the group $\Gamma$ will either be $\mathbb{Z}_n$ or $\mathbb{Z}$, depending on whether the original manifold was compact or not. For the quotient of a compact group to be globally well-defined, the group of identifications has to be finite.

Given the heat kernel, we can then compute the one-loop determinant of the spin-$s$ Laplacian including a mass term by taking the coincident limit $y=x$, integrating over the manifold, and then integrating over the parameter $t$,
\begin{equation}\label{eq:1ldet}
- \frac{1}{2} \ln  \det \left(  -\Delta_{(s)} + M^2  \right) = \frac{1}{2} \int_0^\infty \frac{\, d t}{t} \int_{G/\Gamma} d\mu(x) \  K^{(s)}(x,x;t) \ e^{-M^2  t},
\end{equation}
where we have accounted for the mass $M$ by inserting the exponential inside the integral. The integral over a non-compact space like \AdS{3} will be divergent, but this can be regulated and the divergent piece can be subtracted off by local counterterms.

We now apply the procedure from above to our space of interest, $\asth$. We start by quoting the results for the two factors and then explain how they can be combined. Starting with $\mathbf{S}^3$, the sum in~\eqref{eq:hk} runs over representations labeled by two $\sut$ indices $(\ell,\bar{\ell})$. Since the eigenfunctions are known, the heat kernel on $\mathbf{S}^3$ and its quotient can be calculated as in \cite{David:2009xg}. For the integrated heat kernel the result looks like~\cite{David:2009xg}: 
\begin{equation}\label{eq:hkS3}
        \int_{\mathbf{S}^3/\mathbb{Z}_n} d\mu(x) \  K^{}_{\mathbf{S}^3/\mathbb{Z}_n}(x,x;t) = 
        \sum_{m\in \mathbb{Z}_n} \sum_{\ell, \bar{\ell}}  \chi_{\ell}(z^m) \, \chi_{\bar{\ell}}(\bar{z}^m) \ e^{-C(\ell,\bar{\ell})t} \, ,
\end{equation} 
where $\nu$ and $\bar{\nu}$ are the periods from~\eqref{eq:6Dident}, $\chi_\ell(\nu)$ are spin-$\ell$ $\sut$ characters defined in~\eqref{eq:su2char}, and $C(\ell,\bar{\ell})=2(C_2(\ell)+C_2(\bar{\ell}))-C_2(|\ell-\bar{\ell}|)$ is given by the $\sut$ Casimir's of the two representations $C_2(\ell)=\ell(\ell+1)$. The sum over $m$ reflects the sum over the transformations $\gamma$. The authors of~\cite{David:2009xg} restrict the sum in~\eqref{eq:hk} to the set of transverse and traceless representations, with the goal of eventually performing analytic continuation to get the \AdS{3} result. Since our plan is to dimensionally reduce the SUGRA theory from $AdS_3 \times \mathbf{S}^3$ down to $AdS_3$, we consider $\mathbf{S}^3$ to be an internal space and neglect the dynamics therein. Hence, we drop the sum over representations $(\ell, \bar{\ell})$ and the exponential weight related to the Casimir energy. Overall we are left with an expression summing over the states of a given $\sut$ representation $(\ell,\bar{\ell})$.

The heat kernel on thermal \AdS{3} was computed by analytically continuing the result for $\mathbf{S}^3$. The representations carry two labels, the spin $s$, or rather the helicity, and a continuous label $\lambda$ that gets summed (integrated) over as instructed by~\eqref{eq:hk}.~\cite{David:2009xg} found the integrated heat kernel on \AdS{3} to be\footnote{
    In all of the following equations where we give the integrated heat kernel we will mean integrated only over the manifold as in~\eqref{eq:hkS3}, and we will avoid writing the integral on the left hand side to simplify the expressions. This leaves the $t$ integral to be performed later.
}
\begin{equation}\label{eq:hkAdS3}
 \frac{\tau_2}{2\pi}
\sum_{m\in\mathbb{Z}} \int_0^{\infty} d\lambda 
\left[ \hat{\chi}^{}_{(j_1,j_2)} \left( e^{i \pi m\tau} \right) +
\hat{\chi}_{(j_2,j_1)} \left( e^{i \pi m\tau} \right) \right] \,
e^{-(\lambda^2+s+1)t}
\end{equation}
where 
\begin{equation}
    \hat{\chi}_{j_1,j_2}(\alpha) = \frac{\alpha^{2j_1+1}\bar{\alpha}^{2j_2+1} + \alpha^{-2j_1-1}\bar{\alpha}^{-2j_2-1}}{|\alpha-\alpha^{-1}|^2}
\end{equation}
is the character of the $\mathfrak{sl}(2,\mathbb{C})$ representation with $j_1=\frac{1}{2}(s-1+i\lambda)$ and $j_2=\frac{1}{2}(-s-1+i\lambda)$. The two terms correspond to the two helicities. The sum over $m$ is now infinite, owing to the fact that \AdS{3} is a non-compact manifold.

Since thermal $\asth$ does not globally factorize - $\mathbf{S}^3$ is non-trivially fibered over \AdS{3} as a result of the boundary conditions~\eqref{eq:6Dident} - we cannot simply multiply~\eqref{eq:hkS3} and~\eqref{eq:hkAdS3} to obtain the integrated heat kernel on thermal $\asth$. Instead, by allowing the identifications to go over the entire set~\eqref{eq:6Dident}, we find that the summands of equations~\eqref{eq:hkS3} and~\eqref{eq:hkAdS3} get multiplied, but the sum over images runs over only one set of identifications:
\begin{equation}\label{eq:hk6D}
    \begin{split} 
        \frac{\tau_2}{2\pi}
        \sum_{m\in\mathbb{Z}} \int_0^{\infty} d\lambda
        \left[
        \hat{\chi}^{}_{(j_1,j_2)} \left( e^{i \pi m\tau}  \right) \,
        \chi_{\ell}(z^m) \ \chi_{\bar{\ell}}(\bar{z}^m) +
        \hat{\chi}_{(j_2,j_1)} \left( e^{i \pi m\tau} \right) \,
        \chi_{\bar{\ell}}(z^m) \ \chi_{\ell}(\bar{z}^m)
        \right] \,
        e^{-(\lambda^2 + s+1 + M^2)t} \, ,
    \end{split}
\end{equation}
where we have also included a mass term. The expression above makes evident that each term accounts for a different helicity. Focusing on one helicity, we can perform the Gaussian $\lambda$ integral to find
\begin{equation}\label{eq:hk6D1hel}
    \sum_{m=1}^\infty
    \frac
    {\pi \tau_2 e^{2 \pi i s m \tau_1}  \chi_{\ell}(z^m) \ \chi_{\bar{\ell}}(\bar{z}^m)}
    {\sqrt{4 \pi t} \Big|\sin \pi m\tau\Big|^2} \
    e^{- \frac{4 \pi^2 m^2 \tau_2^2}{4t}}
    e^{-(s+1 + M^2)t} \, .
\end{equation}
The $m=0$ term is missing, because it corresponds to a divergent integral over the volume of \AdS{3}, which can be regulated and removed. Once integrated over $t$ as instructed by~\eqref{eq:1ldet}, this gives the one-loop determinant of the spin-$s$, mass-$M$ Laplacian in \AdS{3} for a field with quantum numbers $(\ell,\bar{\ell})$, whose physical interpretation in \AdS{3} will be explained at the end of this section.

The one-loop partition function can be obtained by exponentiating the one-loop determinant~\eqref{eq:1ldet}, which gives
\begin{equation}\label{eq:Z1l}
Z_{\text{1-loop}} = \text{det}^{-1/2} \left( -\Delta_{(s)}+M^2 \right) \, .
\end{equation}
This directly works for massless scalar, spin-$1/2$ fields and massive fields regardless of the spin, while higher-spin massless fields require gauge-fixing, which introduces ghosts into the action, which then give additional factors of the one-loop determinant in~\eqref{eq:Z1l}.

We will illustrate this using the simplest (in terms of $\sut$ content) example from our spectrum~\eqref{eq:gravqnum}, the massless gravitino with $\left( \ell=\frac{1}{2},\bar{\ell}=0 \right)$. The author of~\cite{David:2009xg} showed that the gauge-fixing in this case produces a spin-$1/2$ ghost, and the partition function is given by
\begin{equation}\label{eq:Z3/2def}
    \text{ln}\,Z^{(3/2)}_{\text{1-loop}} = \frac{1}{2}\text{ln}\,\text{det} \left( -\Delta_{(3/2)}+\frac{9}{4} \right) + \frac{1}{2} \text{ln}\, \text{det} \left( -\Delta_{(1/2)}-\frac{3}{4} \right).
\end{equation}
The mass term in both Laplacians comes from \AdS{3} curvature (for further discussion we refer the reader to~\cref{sec:hkapp}). We can now use the heat kernel on $\asth$ we found~\eqref{eq:hk6D} to compute the one-loop determinants in this equation, by writing them in terms of the heat kernel using~\eqref{eq:1ldet}, and then performing the $t$-integral using~\cite{David:2009xg}
\begin{equation}
    \frac{1}{4\pi^{1/2}}\int_0^{\infty}\frac{dt}{t^{3/2}}e^{-\frac{\alpha^2}{4t}-\beta^2t}=\frac{1}{2\alpha}e^{-\alpha\beta}.
\end{equation}
We find the following expression
\begin{equation}\label{eq:Z3/2onehel}
    \text{ln}\,Z_{\text{one helicity}}^{(3/2)} =
    -\sum_{m=1}^{\infty}
    \frac{(-1)^m}{m}
    \frac{q^{\frac{3m}{2}}}{1-q^m}
    \left( z^{+1/2}+z^{-1/2} \right) \, ,
\end{equation}
where the last factor comes from simplifying $\chi_\frac{1}{2}(z) \, \chi_0(\bar{z})$. After expanding the denominator $(1-q^m)$ in a geometric series, and summing over $m$, we find
\begin{equation}\label{eq:Z3/2resum}
    \text{ln}\,Z_{\text{one helicity}}^{(3/2)} =
    -\sum_{n=2}^{\infty}
    \Big[
    \text{ln} \left( 1-z^{+1/2}\,q^{n-1/2} \right) + \,
    \text{ln} \left( 1-z^{-1/2}\,q^{n-1/2} \right)
    \Big]\, ,
\end{equation}
Note that this is only two of the four gravitini that were mentioned in~\eqref{eq:gravqnum}. Adding the other helicity leads the full gravitino contribution to the partition function
\begin{equation}\label{eq:Z3/2fin}
    Z^{(3/2)} =
    \prod_{n=2}^{\infty} \,
    \Big|1-z^{+1/2}\,q^{n-1/2}\Big|^2 \,
    \Big|1-z^{-1/2}\,q^{n-1/2}\Big|^2 \, .
\end{equation}
Setting $\nu=0$, or $z=1$, we reproduce the result of~\cite{David:2009xg} (or rather the square of their result, see the end of this subsection). The partition functions for all other fields that we will need have been assembled in~\cref{sec:hkapp}.

An alternative way of obtaining the partition function of a field is to perform the following procedure, which we will refer to as multi-particling. First, we write down the partition function of the 1-particle Hilbert space associated with that field. Focusing on one helicity, the states in this space can be obtained from a highest weight state, by applying $L_{-1}$ on it. Depending on the field, the Boltzmann factor associated with the highest weight state has the form of one of the terms in~\eqref{eq:gravchar}, $q^h \, \bar{q}^{\bar{h}} \, \chi_{\ell}(z) \, \chi_{\bar{\ell}}(\bar{z})$, with $q$ term coming from $e^{-\beta E}$, and the $z$ terms associated with chemical potential. The partition function of the whole 1-particle Hilbert space, of say one helicity of the gravitino, is
\begin{equation}\label{eq:Zonepart}
    Z_1(q,z) = \frac{q^{3/2}}{1-q}\,\chi_{1/2}(z) \,.
\end{equation}
Then, to obtain the partition function associated with the multi-particle Fock space, we perform the following operation (see e.g.~\cite{Maldacena:2001b})
\begin{equation}\label{eq:Zonetomulti}
    Z_1(q) \rightarrow \sum_{m=1}^\infty \frac{1}{m}Z_1(q^m) \, .
\end{equation}
Applying this operation to~\eqref{eq:Zonepart} exactly gives~\eqref{eq:Z3/2onehel}, which then, through the same steps as before, gives~\eqref{eq:Z3/2resum}. To obtain the full expression, we repeat this for the other helicity. This method of calculating the partition function relies on the Virasoro boundary algebra, while the heat kernel method is more in the spirit of a pure supergravity calculation, which is what our goal is. We still mention the multi-particling procedure as a quicker way of obtaining the same result, which would come in handy if we considered fields at higher levels of the supergravity spectrum.

Before closing this subsection, we point out that our result for the partition function~\eqref{eq:Z3/2fin} could have been obtained purely by working in \AdS{3}. As we will explain in the next subsection, dimensional reduction on $\mathbf{S}^3$ gives rise to two background gauge fields on \AdS{3}. The fields that fluctuate in this background can be charged under these gauge fields, and carry quantum numbers $(\ell,\bar{\ell})$ specifying the multiplet they belong to. The boundary values of these gauge fields are determined by $\nu$ and $\bar{\nu}$, which can then be viewed as the thermodynamic chemical potentials associated with these charges (for a nice discussion, see~\cite{Kraus:2006a}). To find the heat kernel we would have to solve the \AdS{3} eigenvalue problem accounting for the coupling between the fluctuating fields and background gauge fields. This would amount to adding energy shifts proportional to the chemical potentials $\nu$ and $\bar{\nu}$ to the Laplacian operator, which would ultimately lead to the extra terms exponential in $\nu,\bar{\nu}$ in the expression for the heat kernel~\eqref{eq:hk6D}. By working in 6D, we have avoided solving this eigenvalue problem, and were able to quote the known results for the two factors, and put them together. The main purpose of this paragraph is to point out the three-dimensional gauge field interpretation of our result. Finally, note that from the \AdS{3} perspective,~\eqref{eq:hk6D} is actually the heat kernel for multiple \AdS{3} fields belonging to the same multiplet - for example, when we take the $z=1$ limit in~\eqref{eq:Z3/2fin}, we see that it actually contains two gravitino partition functions.

\subsection{Spectral flow as a sum over saddles}\label{sec:sf}

The results of the previous subsection allow us to build the full one-loop partition function by putting together all of the fields contained in the supergravity spectrum~\eqref{eq:sugraSMK3}. As explained in the last paragraph, this would give us the partition function of fields fluctuating in \AdS{3} with background Chern-Simons gauge fields turned on, with values determined by $\nu$ and $\bar{\nu}$. In this subsection, we will show that choosing the boundary conditions for these gauge fields, $\nu$ and $\bar{\nu}$, corresponds to not one, but an infinite set of configurations, so to obtain the full partition function, we would have to perform a path integral sum over all of these saddles. Furthermore, we will show that different choices of gauge fields correspond to shifts in the values of the boundary conformal algebra charges, which in that context are called spectral flow. This relationship between the spectral flow in the conformal algebra and its bulk interpretation in terms of gauge fields will be key to matching the full partition function to the superconformal characters in the next subsection. While this bulk \AdS{3} interpretation of spectral flow was first pointed out by~\cite{Dijkgraaf:2000}, and expanded on in~\cite{Kraus:2006nb}, we will also be interested in the 6D version of this story in $\asth$, which has been worked out in~\cite{Hansen:2006}.

Spectral flow of a boundary current algebra, which in our case appears as a subalgebra of the $\mathcal{N} = 4$ superconformal algebra, is defined as an automorphism of the algebra, where the generators transform according to
\begin{equation}\label{eq:sfdef}
    \begin{split}
	L_0 & \rightarrow L_0 + 2\, m \ J^3_0 + k \  m^2\, , \\
	J^3_n & \rightarrow J^3_n + \ k \ m \ \delta_{n,0}\, , \\
        J^{\pm}_n & \rightarrow J^{\pm}_{n \pm w}\, ,
    \end{split}
\end{equation}
where, for our purposes, we will take $m$ to be an integer.

Our goal will be to give a bulk interpretation and derivation of~\eqref{eq:sfdef}. To define the boundary charges $L_0$ and $J^3_n$ of equation~\eqref{eq:sfdef} from the bulk, we first need to perform an intermediate step of dimensional reduction from $\asth$ to \AdS{3}. To do this, we take the $\mathbf{S}^3$ part of our metric to be
\begin{equation}\label{eq:S3dimred}
d\theta^2 + \sin^2{\theta} (d\Phi_1 + A_1)^2 + \cos^2{\theta}(d\Phi_2 + A_2)^2.
\end{equation}
This defines two non-Abelian Kaluze-Klein gauge fields on \AdS{3}, which we can further combine into $A$ and $\bar{A}$ to make the $\sut \times\sut$ symmetry of $\mathbf{S}^3$ manifest:
\begin{equation}\label{eq:ALR}
A = A_1 + A_2, \qquad \bar{A} = A_2 - A_1.
\end{equation}
The dynamics of these fields is captured by a Chern-Simons action~\cite{Kraus:2006a,Kraus:2006nb}, which is obtained after integrating over $\mathbf{S}^3$,
\begin{equation}\label{eq:ALact}
    S_{\text{gauge}} =
    - \frac{i k}{4 \pi} \int \ d^3 x
    \ \text{Tr} \left[ A d A + \frac{2}{3} A^3 \right] +
    S_{\text{bnd}} \, ,
\end{equation}
where $k$ is the level of the boundary current algebra. The same equation, but with an opposite sign, holds for $\bar{A}$. Since $A$ will have only one non-zero component, which we take to be the 3-component, we will take $A$ to stand for this sole component for the rest of this subsection.

With the \AdS{3} gauge fields defined, we proceed to explain how the boundary charges are obtained. First, the boundary metric and gauge fields are defined through the standard Fefferman-Graham expansion. Then, functional derivatives are performed on the action with respect to the boundary metric and gauge fields, which define the boundary stress tensor and the currents conjugate to the gauge fields. Requiring the current for $A$ ($\bar{A})$ to be purely left(right)-moving determines the boundary action in~\eqref{eq:ALact}, without which the variational principle would not be well-defined. Finally, the stress tensor and currents are integrated over a circle on the boundary, which results in the corresponding charges. After changing the boundary coordinates to the complex coordinates $w=\phi + i t_E $ and $\bar{w} =\phi - i t_E$, the following equations are obtained
\begin{equation}\label{eq:sfcharg}
        L^{\text{gauge}}_0 = \frac{1}{8 \pi} \oint d w  A_{w}^2 \, , \qquad J_0^3 = i k \oint \frac{d w}{2 \pi i} A_{w} \, .
\end{equation}
An analogous equation holds for $\bar{A}$, with the integral being over $\bar{w}$. Since it turns out that the shifts in $L_0$ corresponding to spectral flow only come from shifts in the gauge field, in \eqref{eq:sfcharg} we have omitted the contribution to $L_0$ coming from the metric.

Following~\cite{Hansen:2006}, we will first explain how spectral flow is generated in the charges from the 6D bulk perspective. To achieve this, consider the following diffeomorphism of $\asth$
\begin{equation}\label{eq:6Ddiffeo}
d\Phi_1 \rightarrow d\Phi_1 + m dw + \bar{m} d\bar{w} \, , \qquad d\Phi_2 \rightarrow d\Phi_2 + m dw - \bar{m} d\bar{w} \, ,
\end{equation}
where we require $m, \, \bar{m} \in \mathbb{Z}$ in order to preserve the periodicities of both $\Phi_1$ and $\Phi_2$. Using equations~\eqref{eq:S3dimred} and~\eqref{eq:ALR} we see that this diffeomorphism induces discrete shifts in the gauge fields living on \AdS{3}
\begin{equation}\label{eq:6DAshft}
A \rightarrow A + 2\, m \, dw \, , \qquad \bar{A} \rightarrow \bar{A} + 2 \, \bar{m} \, d\bar{w} \, .
\end{equation}
In turn, equation~\eqref{eq:sfcharg} implies that these shifts in the gauge fields produce shifts in the boundary charges
\begin{equation}\label{eq:chargshft}
    \begin{split}
	L_0 \rightarrow L_0 + 2m J^3_0 + k m^2\, , \qquad J^3_0 \rightarrow J^3_0 + k m\, ,
    \end{split}
\end{equation}
which exactly correspond to the spectral flow defined in~\eqref{eq:sfdef} with integer $m$. We can then conclude that the bulk diffeomorphism~\eqref{eq:6Ddiffeo} of $\asth$ induces the spectral flow~\eqref{eq:sfdef} in the boundary algebra.

Next, we explain how spectral flow can be argued for purely from an \AdS{3} standpoint, without knowing the higher-dimensional origin of the gauge fields. We start with a theory of Chern-Simons gauge fields $A$ and $\bar{A}$ on \AdS{3}. We can work in a gauge where the radial component of the gauge fields is set to zero, leaving only the $(w,\bar{w})$ components of $A$. Furthermore, we impose boundary conditions, consistent with the choice of the boundary action discussed under \eqref{eq:ALact}, as follows\footnote{
    Our expression is not identical to that of \cite{Kraus:2006nb} as they considered the BTZ black hole background, while we are working in thermal \AdS{3}. The two expressions can easily be obtained one from the other, as these two spaces are the same manifold, and essentially differ by a coordinate transformation.
}
%
%
\begin{equation}\label{eq:Abndcnd}
    A_{\bar{w}} \propto \frac{\nu}{\tau} \, ,
    \qquad
    \bar{A}_{w} \propto \frac{ \bar{\nu}}{\tau} \, .
\end{equation}
This choice will be further justified in the next subsection by comparison to the superconformal character. This leaves one component for each gauge field unspecified by the boundary conditions, which is consistent with the fact that the gauge fields also have to satisfy an additional holonomy constraint, which requires a Wilson line for a charged particle around a contractable boundary cycle to satisfy
\begin{equation}\label{eq:holcnstr}
    \exp \left[ {i \int_{\gamma} A} \right]= -1.
\end{equation}
This condition is equivalent to imposing that $A$ and $\bar{A}$ be non-singular. Solving this constraint determines the remaining component of each gauge field
\begin{equation}\label{eq:3DAshft}
    A_{w} \propto - \frac{\nu}{\tau} + 2 m \, ,
    \qquad
    \bar{A}_{w} \propto - \frac{\bar{\nu}}{\tau} + 2 \bar{m}\,,
\end{equation}
with $m,\, \bar{m} \in \mathbb{Z}$. This corresponds precisely to the gauge configurations in equation~\eqref{eq:6DAshft} generated by the 6D diffeomorphism, which as we already saw, correspond to spectral flow in the boundary charges.

This analysis shows us that when we fix the boundary conditions to compute the partition function path integral, they single out a whole set of gauge field configurations, and the partition function has to be computed, using the methods of \cref{sec:hkgeneral}, in each of these backgrounds, which we will refer to as the ($m,\bar{m}$) spectral flow sectors, given their relationship to the algebra spectral flow. Once these are computed, their contributions are summed up as required by the path integral sum over saddles.

\subsection{The small \texorpdfstring{$\mathcal{N}=4$}{N4} characters from supergravity}\label{sec:K3char}

We now combine the results of the previous subsections to find the one-loop partition function at level $j=0$ of the supergravity spectrum \eqref{eq:sugraSMK3}, which, as we saw, only includes the supergraviton multiplet. We proceed in two steps. First, in the $(m=0,\bar{m}=0)$ spectral flow sector, we put together the partition functions of all fields of this multiplet $\mathrm{Z}_{(0,0)}$. Then, we compute the partition function in the $(m,\bar{m})$ sector $\mathrm{Z}_{(m,\bar{m})}$, by performing the gauge field shifts, and we sum over sectors. Once we have the full partition function, we compare it to the product of the left and right superconformal characters \eqref{eq:sm4charorigin} of the short representations introduced at the beginning of this section, and we show that they match.

To find the one-loop partition function at $(m=0,\bar{m}=0)$, we multiply the contribution of each field in~\eqref{eq:gravqnum}, which have been computed by the methods of \cref{sec:hkgeneral} and compiled in \cref{sec:hkapp}. We find
\begin{equation}\label{eq:Zseed}
    \begin{split}
        & \mathrm{Z}_{(0,0)}^{\text{1-loop}} = 
        \underbrace{\frac{1}{\prod_{n=1}^\infty
        \Big| 1 - z^{+1} \, q^n \Big|^2
        \Big| 1 - q^n \Big|^2
        \Big| 1 - z^{-1} \, q^n \Big|^2}
        }_{\text{3}\text{ vectors}}
        \times \\ &
        \ \ \underbrace{\prod_{n=2}^\infty
        \Big|1-z^{1/2}\,q^{n-1/2}\Big|^4 \
        \Big|1-z^{-1/2}\,q^{n-1/2}\Big|^4}
        _{\text{4 gravitini}}
        \times
        \underbrace{\frac{1}{\prod_{n=2}^\infty
        \Big| 1-q^n \Big|^2}}_{\text{graviton}} \, .
    \end{split}
\end{equation}

If we wanted the whole partition function (up to one-loop), we would also need to add the semiclassical part. This can be found by exponentiating the \AdS{3} on-shell action in the presence of gauge fields, which was obtained in~\cite{Kraus:2006a,Kraus:2006nb}
\begin{equation}\label{eq:Sonshll}
    S_{\text{on-shell}} =
    -\frac{i\pi k\tau}{2}
    +\frac{i\pi \bar{k}\bar{\tau}}{2} -\pi\tau_2(A_{\bar{w}}^2 +\bar{A}_w^2).
\end{equation}
The first two terms give the well-known $|q|^{-\frac{c}{12}}$ semiclassical partition function of \AdS{3} gravity, while the remaining terms give the gauge field contributions. We will leave out both of these in our calculations, for a cleaner comparison to the superconformal character.

To find the contribution of the $(m,\bar{m})$ spectral flow sector, we need to re-evaluate the partition function \eqref{eq:Zseed} in the shifted gauge field configuration of equation \eqref{eq:3DAshft}, but this is equivalent to shifts in the modular parameter $\nu\rightarrow\nu+2m\tau$, which corresponds to replacing $z\rightarrow z \, q^{2m}$. Furthermore, using \eqref{eq:Sonshll}, we see that there are additional contributions $q^{km^2} \, z^{km}$ to each sector coming from the semiclassical action. Putting this together and summing over the sectors, we find the full one-loop partition function
\begin{equation}\label{eq:Zsf}
    \begin{split}
        \mathrm{Z}_{\text{SUGRA}}^{\text{1-loop}} =
        &\sum_{m,\bar{m}=0}^{\infty} 
        q^{km^2+k\bar{m}^2} z^{km+k\bar{m}} \\ &
        \times \,
        \frac
        {\prod_{n=2}^\infty \,
        \Big|1-z^{1/2} \,q^{m+n-1/2}\Big|^4 \,
        \Big|1-z^{-1/2}\,q^{-m+n-1/2}\Big|^4}
        {\prod_{n=1}^\infty \,
        \Big| 1 - z^{}\, q^{2m+n} \Big|^2 \,
        \Big| 1 - q^n \Big|^2 \,
        \Big| 1 - z^{-1} \, q^{-2m+n} \Big|^2 \,
        \prod_{n=2}^\infty \,
        \Big| 1-q^n \Big|^2} \, .
    \end{split}
\end{equation}

We now turn to the superconformal character of the $\mathcal{N}=4$ superconformal algebra.
We should compare our partition function to the vacuum ${\ell}=0$ holomorphic and ${\bar{\ell}}=0$ antiholomorphic short representation characters, which are the $(\ell,\bar{\ell})$ $\sut$ labels of the graviton in the supergraviton multiplet. We will first show that the $m=0$ term of the sum in the character \eqref{eq:sm4charorigin} matches (the holomorphic part of) \eqref{eq:Zseed}. The $m=0$ term of \eqref{eq:sm4charorigin} reads\footnote{
    Since we are interested in the trace \eqref{eq:sm4chardef} with the $(-1)^F$ insertion, we set $\delta=1$ in \eqref{eq:sm4charorigin}.}
\begin{equation}\label{eq:charseed}
    \begin{split}
        \schr{}(0)\Big|_{m=0} =
        \prod_{n=1}^\infty
        \frac
        {\Big(1-z^{+1/2}\,q^{n-1/2}\Big)^2 \,
        \Big(1-z^{-1/2}\,q^{n-1/2}\Big)^2}
        {\Big( 1 - z \, q^{n} \Big) \,
        \Big( 1 - q^n \Big)^2 \,
        \Big( 1 - z^{-1} \, q^{n-1} \Big)} \,
        \left[
        \frac{1}
        {\Big(1-z^{+1/2}\,q^{-1/2}\Big)^2} - \frac{z^{-1}}
        {\Big(1-z^{-1/2}\,q^{-1/2}\Big)^2}
        \right] \, .
    \end{split}
\end{equation}
The second factor in this equation simplifies to
\begin{equation}\label{eq:charalg}
    \frac{(1-z^{-1}) \,
    (1-q)}
    {(1-z^{+1/2}\,q^{1/2})^2 \,
    (1-z^{-1/2}\,q^{1/2})^2} \, ,
\end{equation}
which exactly cancels the right pieces in the first factor of \eqref{eq:charseed} to give the holomorphic part of \eqref{eq:Zseed} (after shifting some of the infinite products). The ${\bar{\schr{}}(0)}\Big|_{m=0}$ would match the anti-holomorphic part of \eqref{eq:Zseed}. So, the partition function and character agree in the zero spectral flow sector.

To perform the rest of the comparison, we can exploit spectral flow in the conformal algebra to reinterpret the full character as being generated from the $m=0$ seed element \eqref{eq:charseed} as follows. Given the definition of the character \eqref{eq:sm4chardef} and spectral flow \eqref{eq:sfdef}, if we label $\mathfrak{f}(q,z) = \schr{}(0)\Big|_{m=0}$, performing spectral flow and summing over the sectors amounts to the following operation
\begin{equation}\label{eq:sfonchar}
    \mathfrak{f}(q,z) \rightarrow \sum_{m\in \mathbb{Z}}\, q^{k\,m^2}\, z^{k\,m}\, \mathfrak{f}(q,z \, q^{2m}) \,.
\end{equation}
To show that this recovers the full character, we will start from the $m=0$ term of the character written in terms of theta functions \eqref{eq:sm4chartheta}
\begin{equation}\label{eq:charseedtheta}
\begin{split}
    \schr{}(0)\Big|_{m=0} =
    q^{\frac{1}{4}} \,
    \frac
    {\vartheta_{4}^2 \left( z^{1/2}, q \right)}
    {\eta(q)^3 \, \vartheta_1 (z, q)} 
    \left[ \,
    \mu(q,z) - \mu(q,z^{-1})\,
    \right]_{m=0} \, .
\end{split}
\end{equation} 
One can check that applying $\mathfrak{f}(q,z)\rightarrow \mathfrak{f}(q,z\,q^m)$ to the first factor in this equation, and using the theta function modular transformation equation $\vartheta(q,z \,q^{2m}) = q^{-2m^2} \, z^{-2m} \ \vartheta(q,z)$, gives
\begin{equation}\label{eq:thetashft}
\begin{split}
    \frac
    {\vartheta_{4}^2 \left( z^{1/2}, q \right)}
    {\eta(q)^3 \, \vartheta_1 (z, q)}
    \rightarrow
    q^{-m^2} \, z^{-m} \,
    \frac
    {\vartheta_{3}^2 \left( z^{1/2}, q \right)}
    {\eta(q)^3 \, \vartheta_1 (z, q)}\, .
\end{split}
\end{equation}
For the second factor in \eqref{eq:charseedtheta}, one can also check that performing \eqref{eq:sfonchar} gives
\begin{equation}\label{eq:mushft}
\begin{split}
    \left[ \,
    \mu(q,z) - \mu(q,z^{-1})\,
    \right]_{m=0}
    \rightarrow
    q^{m^2} z^{m}
    \left[ \,
    \mu(q,z) - \mu(q,z^{-1})\,
    \right]_m \, .
\end{split}
\end{equation}
Note that the sign of $m$ has to be flipped in the second $\mu_m$ term, which we get to do because in the end we perform a sum over all integers. Putting together~\eqref{eq:thetashft} and~\eqref{eq:mushft}, adding the prefactor from \eqref{eq:sfonchar}, and performing the sum, shows that spectral flow performed on the seed element \eqref{eq:charseedtheta} gives exactly the full superconformal character
\begin{equation}\label{eq:fullchartheta}
\begin{split}
    \schr{}(0) =
    q^{\frac{1}{4}} \, 
    \frac{\vartheta_{4}^2 ( z^{1/2}, q)}{\eta(q)^3 \, \vartheta_1 (z, q)} 
    \left[ \,
    \mu(q,z) - \mu(q,z^{-1})\,
    \right]\,.
\end{split}
\end{equation} 

Putting everything together, we have shown that the full one-loop supergravity partition function and the superconformal character can both be written as sums over spectral flow sectors. Furthermore, we have shown that in the $(m=0,\bar{m}=0)$ sector these two objects agree. By the arguments of \cref{sec:sf}, which established the equivalency of bulk and boundary spectral flow, this shows that the one-loop partition function and \emph{small} $\mathcal{N}=4$ superconformal characters agree sector by sector
\begin{equation}\label{eq:agree}
    \mathrm{Z}_{\text{SUGRA}}^{\text{1-loop}} =
    \schr{}(0) \,\bar{\schr{}}(0) \, ,
\end{equation}
confirming the predictions of \cite{Maloney:2007ud}. The computation for higher levels $j\geq1/2$ would involve additional multiplets and fields with higher quantum numbers, but would proceed analogously to the one outlined here, in a straightforward manner.

\subsection{The case of \texorpdfstring{$\asx{T^4}$}{AdS3T4}}\label{sec:T4onel}

We now turn to the $T^4$ compactification. In this case, when we compactify from 10D to 6D, there are additional massless gravitini, leading to a 6D $\mathcal{N} = (2,2)$ supergravity theory. However, further compactification on $\asth$ does not preserve this enhancement, but only leads to a 3D theory with $\psut \times \overline{\psut}$ superisometry group. The resulting spectrum can be obtained by combining the analysis of~\cite{Deger:1998nm} and~\cite{Larsen:1998xm}.
In terms of the notation used in~\eqref{eq:sugraSMK3} we can write the spectrum as 
\begin{equation}\label{eq:sugraSMT4}
\begin{split}
\sccft^{(2)}(0)  +  2\,\sccft^{(\frac{3}{2})}(0) + 
\sum_{j\geq \frac{1}{2}}  \left[ \sccft^{(2)}(j) + \mathbf{4}_5\, \sccft^{(\frac{3}{2})}(j) +   \mathbf{5}_5\,  \sccft^{(1)}(j) + \sccft^{(1)} \left( j+ \frac{1}{2} \right)  \right]  + \text{cc} \,.
\end{split}
\end{equation}
We have again singled out the boundary supergravitons, which now not only comprise a helicity-$2$ multiplet (and its conjugate), but also a helicity-$\frac{3}{2}$ multiplet that arises from the additional symmetries preserved by $T^4$. The latter multiplet contains the four $\mathrm{u}(1)$ currents, and their spin-$\frac{1}{2}$ partners, with the latter originating from the additional supersymmetries preserved by the compactification to 6D.

We can find the field content and corresponding $\sut$ charges of the $j=0$ level by repeating the analysis of \cref{sec:K3spec}. The graviton multiplet $\sccft^{(2)}(0)$ remains unchanged. Using \eqref{eq:psu2chartrunc} and \eqref{eq:helchar} we find that the $\frac{3}{2}-$multiplet takes the form
\begin{equation}\label{eq:3/2char}
    \sccft^{(\frac{3}{2})}(0) =  q^{1/2} \chi_{1/2}(z) - 2 \, q.
\end{equation}
Therefore, the corresponding field content comprises a massless spin-$\frac{1}{2}$ fermion that transforms as an $\sut$ doublet and two massless vectors:
\begin{equation}\label{eq:3/2qnum}
    1\times \left( \frac{1}{2}, 0,\frac{1}{2},0 \right)
    +
    2\,  \times (1,0,0,0) \, .
\end{equation}

The one-loop partition function is the same as that of the $K3$ compactification \eqref{eq:Zseed} supplemented with the contributions coming from the additional fields we found above, but squared, since there are two $\sccft^{(\frac{3}{2})}(0)$ multiplets in the spectrum \eqref{eq:sugraSMT4}:
\begin{equation}\label{eq:ZT4extra}
    \underbrace{\prod_{n=1}^\infty
        \Big|1-z^{+1/2}\,q^{n-1/2}\Big|^4 \
        \Big|1-z^{-1/2}\,q^{n-1/2}\Big|^4}
        _{\text{4 spin-}\frac{1}{2}\text{ fermions}}
        \times
        \underbrace{\frac{1}{\prod_{n=1}^\infty
        \Big| 1 - q^n \Big|^8}
        }_{\text{4}\text{ vectors}} \, .
\end{equation}
The matching to boundary algebra characters for this multiplet is much more straightforward. The holomorphic piece of \eqref{eq:ZT4extra} can be directly matched onto the character of four free $\mathrm{u}(1)$ currents and their four superpartner spin-$1/2$ fermionic currents
\begin{equation}\label{eq:T4bndchar}
    \prod_{n=1}^{\infty} \, \frac
    {\Big(1-z^{+1/2}\,q^{n-1/2}\Big)^2\,
    \Big(1-z^{-1/2}\,q^{n-1/2}\Big)^2}
    {\Big(1-q^n\Big)^4} \, .
\end{equation}

\section{One-loop supergravity determinant around \texorpdfstring{$\asx{\stso}$}{AdS3S3S1}}\label{sec:S3S1onel}

The analysis of $\asx{\mathbf{S}^3\times\mathbf{S}^1}$ will follow as a straightforward generalization of the previous section, with the main difference appearing at the level of the isometry and asymptotic algebras. $\asx{\mathbf{S}^3\times\mathbf{S}^1}$ is a Type II background\footnote{
    We won't distinguish between IIA and IIB, as they are related by T-duality on the $\mathbf{S}^1$.
}. It can be supported by either NS-NS or R-R two form flux, or a combination thereof, and it preserves 8 supercharges.

The superisometry group of $\asx{\mathbf{S}^3\times\mathbf{S}^1}$ consists of two copies of $\text{D}(2,1|\alpha)\times\overline{\text{D}(2,1|\alpha)}$, with $\alpha=R_+^2/R_-^2$ being a continuous parameter determined by the radii of the two $\mathbf{S}^3$ factors, $R_\pm$. The algebra is generated by $\{L_0,L_{\pm1},J_0^{\pm,i},G_{\pm\frac{1}{2}}^a\}$, with $i=1,2,3$ and $a=1,2$. The commutation relations can be found in~\cref{sec:largealg}. The isometries of the two $\mathbf{S}^3$ factors now result in two $\sut_\pm$ bosonic subalgebras, in addition to the $\slt$ associated with the symmetries of \AdS{3}.

A representation of $\text{D}(2,1|\alpha)$ comprises of states obtained by acting with the raising operators on a highest weight state $\ket{h,j_\pm}$, which is now labeled by two $\sut$ spins in addition to the weight. Furthermore, unitary representations satisfy the following bound
\begin{equation}\label{eq:largebndiso}
    h \geq \left(\frac{1}{1 + \alpha} j_{-} + \frac{\alpha}{1 + \alpha} j_{+} \right) \, .
\end{equation}
The bound is saturated by the short representations. Generically supergravity only sees short representations, as was the case with the compactifications on $T^4$ and $K3$, but in this section we will also have long representations appearing in the spectrum. The representation content will be presented in~\cref{sec:S3S1spec} through the character.

The fact that this background involves two $\mathbf{S}^3$ factors results in a more complex asymptotic algebra different from the one we had in the previous section, the \emph{large} $\mathcal{N}=4$ algebra $\mathcal{A}_{\gamma}$, which depends on the relative size of the two spheres through the continuous parameter $\gamma=\frac{\alpha}{1+\alpha}$. In addition to the Virasoro generators, two sets of $\sut$ generators, and their superpartners, it also includes $\mathrm{u}(1)$ generators and their superpartners, $\{L_n, J^{\pm,i}_n, G_r^a, U_n,Q^a_r \}$, where $n \in \mathbb{Z}$, $r \in \mathbb{Z}+\frac{1}{2}$ or $r\in \mathbb{Z}$ in the Neveu-Schwarz and Ramond sectors, respectively, and $i \in \{1,2,3\}$ and $a\in \{0,1,2,3\}$ are $\sut_\pm$ vector and bispinor indices. The central charge is $c=\frac{3\,\lads}{2\,G_N}$, while the levels of the two $\sut_\pm$ current algebras $k_\pm$ are given by \eqref{eq:lg4cgamma}. The commutation relations can also be found in~\cref{sec:largealg}.

Representations are built atop of the highest weight state $\ket{h,\ell_\pm}$. We will again focus on representations in the NS sector, so $r\in \mathbb{Z} +\frac{1}{2}$. Unitary representations are constrained by having the $\sut_\pm$ spins satisfy the following bound
\begin{equation}\label{eq:largebndasy}
    h \geq \frac{1}{k_++k_-}\left[
    k_+\ell_-+k_-\ell_++(\ell_+-\ell_-)^2+u^2
    \right] \, .
\end{equation}
States saturating the bound are short BPS multiplets, and their content is summed up by the character formulas given in~\cref{sec:largealg}.

One can explicitly check in the commutation relations~\eqref{eq:lg4comm2} that $\{L_n, J^{\pm,i}_n, G_r^a\}$ do not decouple from the rest of the algebra $\{U_n,Q^a_r\}$, as was the case for the isometry group. By performing a field redefinition \cite{Goddard:1988}, we can decouple the $\{U_n,Q^a_r \}$ from the rest of ${\mathcal{A}}_{\gamma}$, which results in two subalgebra factors. The $\{L_n, J^{\pm,i}_n, G_r^a\}$ generators form the subalgebra $\widetilde{\mathcal{A}}_{\gamma}$ with central charge $c-3$, while $\{U_n,Q^a_r \}$ generates the subalgebra $\mathcal{A}_\text{QU}$, which has a free field realization. As a result, the superconformal characters of ${\mathcal{A}}_{\gamma}$ factorize, as can be seen in \cref{sec:largealg}.

\subsection{$\text{D}(2,1|\alpha)$ and the spectrum}\label{sec:S3S1spec}

We will present the three-dimensional spectrum in terms of characters of $\text{D}(2,1|\alpha) \times \overline{\text{D}(2,1|\alpha)}$. We define these characters as follows
\begin{equation}\label{eq:Dchardef}
    \schr{}^\text{g}(h,j_+,j_-) =
    \Tr_{_\text{}}\left((-1)^{F} \, 
    q^{L_0} \, z_+^{J_0^{+,3}} \,  \, z_-^{J_0^{-,3}} \right)\, .
\end{equation}
We will focus on the characters for the short representations, as those will be needed for the part of the spectrum which we will work with. Applying the raising operators to the highest weight state in a short representation results in eight states, which we organize in terms of the weight as
\begin{equation}\label{eq:DcharS}
    \begin{split}
        \schr{\text{S}}^\text{g}(\ell_+,\ell_-)
        & = q^{h} \
        \chi_{\ell_{+}} (z_{+})
        \chi_{\ell_{-}} (z_{-}) \\
        & + q^{h+\tfrac{1}{2}}
        \Big[ \chi_{\ell_{+}+\tfrac{1}{2}} (z_{+}) \chi_{\ell_{-}-\tfrac{1}{2}} (z_{-}) + \chi_{\ell_{+}-\tfrac{1}{2}} (z_{+}) \chi_{\ell_{-}+\tfrac{1}{2}} (z_{-}) + \chi_{\ell_{+}-\tfrac{1}{2}} (z_{+}) \chi_{\ell_{-}-\tfrac{1}{2}} (z_{-}) \Big] \\
        &
        + q^{h+1} \Big[
        \chi_{\ell_{+}} (z_{+})
        \chi_{\ell_{-}-1} (z_{-}) +
        \chi_{\ell_{+}-1} (z_{+})
        \chi_{\ell_{-}} (z_{-}) +
        \chi_{\ell_{+}} (z_{+})
        \chi_{\ell_{-}} (z_{-}) \Big] \\
        & + q^{h+\tfrac{3}{2}} \
        \chi_{l_{+}-\tfrac{1}{2}} (z_{+})
        \chi_{l_{-}-\tfrac{1}{2}} (z_{-}),
    \end{split}
\end{equation}
where $\chi_j(z_\pm)$ are $\sut_\pm$ characters, $z_\pm$ are the elliptic parameters associated with the two $\mathbf{S}^3$ pieces, and the weight saturates \eqref{eq:largebndiso} $h=\frac{\alpha \ell_+ +\ell_-}{1+\alpha} = \gamma \ell_+ + (1-\gamma) \ell_-$. At small values of $\ell_{\pm}$ there are truncations in the character. The lowest non-trivial character has $h=\ell_\pm=\frac{1}{2}$, and is given by
\begin{equation}\label{eq:Dchartrunc}
    \schr{\text{S}}^\text{g} \left( \frac{1}{2},\frac{1}{2} \right) = - q^\frac{1}{2} \chi_{\frac{1}{2}} (z_{+}) \chi_{\frac{1}{2}} (z_{-})
    + q \left[\,1 + \chi_1(z_{+}) + \chi_2(z_{-}) \, \right]
    - q^\frac{3}{2} \chi_{\frac{1}{2}} (z_{+}) \chi_{\frac{1}{2}} (z_{-})
    + q^2.
\end{equation}
The spectrum of supergravity compactified on $\mathbf{S}^3\times\mathbf{S}^3\times\mathbf{S}^1$ was worked out in detail in \cite{Eberhardt:2017fsi} by dimensionally reducing and decomposing the fields into harmonics. When assembled in terms of the $\schr{\text{S}}^\text{g}(\ell_+,\ell_-)$ characters, the spectrum reads
\begin{equation}\label{eq:sugraSMS3S1}
    \sum_{j=0}^\infty \left|
    \schr{\text{S}}^\text{g}(j,j) +
    \schr{\text{S}}^\text{g} \left( j+\frac{1}{2},j+\frac{1}{2} \right)
    \right|^2
    + \text{Long} \, .
\end{equation}
As noted earlier, the spectrum includes both short and long representations. Furthermore, all short representations have $j_+=j_-$. From~\eqref{eq:largebndiso} and~\eqref{eq:largebndasy} we see that this is the condition that the two bounds agree, which is a necessary conditions for the representations of the isometry algebra to be lifted into the representations of the asymptotic algebra. We will rely on this fact when we match the supergravity partition function of the lowest short multiplet to the superconformal characters of the asymptotic algebra. We will focus on the $j=0$ level in \eqref{eq:sugraSMS3S1}. The multiplets can again be characterized by the highest helicity $s=h-\overline{h}$ that appears. We can see that in the BPS part of the spectrum, the multiplets have the structure of $\schr{\text{S}}^\text{g}(j_1,j_1) \times \overline{\schr{\text{S}}^\text{g}(j_2,j_2)}$. One can check using~\eqref{eq:Dchartrunc} that, if $j_1$ and $j_2$ differ by one half, the multiplet is a spin-2 multiplet, and if they're equal, it is a spin-3/2 multiplet. We will ignore $\schr{\text{S}}^\text{g}(0,0) \ \overline{\schr{\text{S}}^\text{g}(0,0)}$ as it only includes one state, the vacuum. We will be interested in the massless part of the spectrum, which sits in the spin-2 multiplet, which in the notation of~\eqref{eq:sugraSMS3S1} is $\schr{\text{S}}^\text{g}(0,0) \ \overline{\schr{\text{S}}^\text{g}(\frac{1}{2},\frac{1}{2})} \ + \  \text{c.c}$. Using equation~\eqref{eq:Dchartrunc}, we can write out the character, which summarizes the field content
\begin{equation}\label{eq:gravcharlarge}
    \left[- q^\frac{1}{2} \chi_{\frac{1}{2}} (z_{+}) \chi_{\frac{1}{2}} (z_{-}) + q \left[ 1 + \chi_1(z_{+}) + \chi_2(z_{-}) \right] - q^\frac{3}{2} \chi_{\frac{1}{2}} (z_{+}) \chi_{\frac{1}{2}} (z_{-}) + q^2 \right] \times \Big[ \bar{q}^{0} \Big].
\end{equation}
Now each term has the form of $q^h \, \bar{q}^{\bar{h}} \, \chi_{\ell_+}(z_+) \, \chi_{\bar{\ell}_+}(\bar{z}_+) \, \chi_{\ell_-}(z_-) \, \chi_{\bar{\ell}_-}(\bar{z}_-)$, which represents a field with a certain helicity, and $\sut_\pm$ quantum numbers. We can extract the field content of the multiplet and the quantum numbers $(h,\bar{h},\ell_+,\bar{\ell}_+,\ell_-,\bar{\ell}_-)$ of each term. We will split up the fields into two parts, with the first being
\begin{equation}\label{eq:gravqnumlarge1}
\begin{split}
    \left( \frac{1}{2},0,\frac{1}{2},0,\frac{1}{2},0 \right) \ + \
    \left( 1,0,0,0,0,0 \right) \, ,
\end{split}
\end{equation}
which consists of an uncharged spin-1 field and a spin-1/2 fermion with mixed $\sut_\pm$ content. In \AdS{3}, these are associated with large gauge transformations of a $\mathrm{u}(1)$ vector field, and corresponding large SUSY transformation. The remaining fields are given by
\begin{equation}\label{eq:gravqnumlarge2}
\begin{split}
    \left( 1,0,1,0,0,0 \right) \ + \
    \left( 1,0,0,0,1,0 \right) \ + \
    \left( \frac{3}{2},0,\frac{1}{2},0,\frac{1}{2},0 \right) \ + \
    \left( 2,0,0,0,0,0 \right)\, ,
\end{split}
\end{equation}
where the first two terms are spin-1 fields which separately transform as $\sut_+$ and $\sut_-$ triplets, then there is a spin-3/2 field with mixed $\sut_\pm$ content, and finally the spin-2 modes. The complex conjugate multiplet gives the opposite multiplicities. This confirms that the massless field structure agrees with the generators of the superisometry algebra.

For the rest of the spectrum, one would have to multiply the characters and find their field content. As far as we know, there are no tables summarizing all multiplets, like there were for the previous compactifications.

\subsection{Heat kernel and spectral flow with multiple gauge fields}\label{sec:S3S1hk}

The analyses of \cref{sec:hkgeneral} and \cref{sec:sf} to backgrounds with more than two gauge fields follow in a straightforward manner. We take the line-element of \AdS{3}$\times \mathbf{S}^3 \times \mathbf{S}^3$ to be
\begin{equation}\label{eq:9Dlelem}
ds^2_{_\text{9D}} =  d\rho^2 + \cosh^2\rho\, dt_\text{E}^2 + R^2_+d\Omega^2_+ + R^2_- d\Omega^2_- \, ,
\end{equation}
with the period of the Euclidean circle and the angles on the two $\mathbf{S}^3$ being set as before. We impose the following twisted boundary conditions
\begin{equation}\label{eq:9Dident}
\begin{split}
&
(\tE, \varphi, \phi_{+,1},\phi_{+,2},\phi_{-,1},\phi_{-,2}) \sim \\
&
(\tE + \beta, \varphi +i\, \beta\, \mu, \phi_{+,1} +i\, \beta\, \nu_{+,1}, \phi_{+,2} +i\, \beta\,\nu_{+,2},\phi_{-,1} +i\, \beta\, \nu_{-,1}, \phi_{-,2} +i\, \beta\,\nu_{-,2}) \,,
\end{split}
\end{equation}
which define the elliptic parameters $\nu_{\pm,1,2}$, which can be further repackaged into
\begin{equation}\label{eq:nupm}
\nu_+ = \beta\, \nu_{+,2} - \beta\, \nu_{+,1} \,  \rightarrow \, z_+ = e^{2\pi i\,\nu_+} \,, \qquad \nu_- = \beta\, \nu_{-,2} - \beta\, \nu_{-,1} \,  \rightarrow \, z_- = e^{2\pi i\,\nu_-} \,.
\end{equation}
The $\nu_{\pm}$ will correspond to complex chemical potentials associated with background gauge fields that will come from dimensional reduction, of which there will now be four.

The additional $\mathbf{S}^3$ and the extended identifications of~\eqref{eq:9Dident} result in additional $\sut$ characters in~\eqref{eq:hk6D}. We quote the result for the integrated heat kernel for one helicity
\begin{equation}\label{eq:hk9D}
    \begin{split}
        \frac{\tau_2}{2\pi}
        \sum_{m\in\mathbb{Z}} \int_o^{\infty} d\lambda
        \left[
        \hat{\chi}^{}_{(j_1,j_2)}(e^\frac{im\tau}{2}) \,
        \chi_{\ell_+}(z^m_+) \ \chi_{\bar{\ell}_+}(\bar{z}^m_+) \
        \chi_{\ell_-}(z^m_-) \ \chi_{\bar{\ell}_-}(\bar{z}^m_-)
        \right] \,
        e^{-(\lambda^2 + s+1 + M^2)t}
    \end{split}
\end{equation}
We can use this to compute the one-loop determinant and partition function of the fields in~\eqref{eq:gravqnumlarge1} and~\eqref{eq:gravqnumlarge2}. As an example, we again present the spin-3/2 fields, as this is the only field that has mixed $\sut_\pm$ content, and therefore requires all of~\eqref{eq:hk9D}. Most of the analysis that follows is identical to that in~\cref{sec:hkgeneral}. We quote the result for one helicity,
\begin{equation}\label{eq:Z3/2onehellarge}
    \text{ln}\,Z^{(3/2)} =
    -\sum_{m=1}^{\infty}
    \frac{(-1)^m}{m}
    \frac{q^{\frac{3m}{2}}}{1-q^m} \,
    (z_+^{1/2}+z_+^{-1/2}) \,
    (z_-^{1/2}+z_-^{-1/2}) \, ,
\end{equation}
where we see that the effect of~\eqref{eq:hk9D} is to add additional $\sut$ characters to the one-loop determinant. Completing the calculation and adding the opposite helicity fields results in
\begin{equation}\label{eq:Z3/2finlarge}
    \begin{split}
    Z^{(3/2)} =
    \prod_{n=2}^{\infty} \, &
    \Big|1-z_+^{1/2}\, z_-^{1/2} \, q^{n-1/2}\Big|^2 \,
    \Big|1-z_+^{-1/2} \, z_-^{1/2} \, q^{n-1/2}\Big|^2 \, \\ &
    \times
    \Big|1-z_+^{1/2} \, z_-^{-1/2} \, q^{n-1/2}\Big|^2 \,
    \Big|1-z_+^{-1/2} \, z_-^{-1/2} \, q^{n-1/2}\Big|^2 \, .
    \end{split}
\end{equation}
which can be rewritten in a more condensed form
\begin{equation}\label{eq:Z3/2largecmprs}
    Z^{(3/2)} =
    \prod_{n=2}^{\infty} \,
    \prod_{\lambda_\pm\in\{\pm\} } \,
    \Big|1-z_+^{\lambda_+/2} \, z_-^{\lambda_-/2} \, q^{n-1/2}\Big|^2 \, .
\end{equation}
This also could have been obtained using the multi-particling technique described in \cref{sec:hkgeneral}.

Turning to spectral flow, we first define the Chern-Simons gauge fields obtained from dimensional reduction of~\eqref{eq:9Dlelem} onto \AdS{3} as follows
\begin{equation}\label{eq:S3pmdimred}
\begin{split}
d\theta_+^2  & + \sin^2{\theta_+} (d\Phi_{+,1} + A_{+,1})^2 + \cos^2{\theta_+}(d\Phi_{+,2} + A_{+,2})^2 \\ & + d\theta_-^2 + \sin^2{\theta_-} (d\Phi_{-,1} + A_{-,1})^2 + \cos^2{\theta_-}(d\Phi_{-,2} + A_{-,2})^2 \, .
\end{split}
\end{equation}
These can further be combined into $A_{\pm}$ and $\bar{A}_{\pm}$:
\begin{equation}\label{eq:ApmLR}
\begin{split}
A_{+} = A_{+,1} + A_{+,2}, \ \ \ \bar{A}_{+} = A_{+,2} - A_{+,1},\\ A_{-} = A_{-,1} + A_{-,2}, \ \ \ \bar{A}_{-} = A_{-,2} - A_{-,1}.
\end{split}
\end{equation}
The boundary charges $L^{gauge}_0$ and $J_0^\pm$ take the same form as before~\eqref{eq:sfcharg}, with the obvious generalization from $J_0$ to $J_0^\pm$.

Spectral flow can either be generated by a diffeomorphism of \AdS{3}$\times \mathbf{S}^3 \times \mathbf{S}^3$ analogous to~\eqref{eq:6Ddiffeo}, which induces integer shifts in the gauge fields, or by working in \AdS{3} and imposing the following boundary conditions on the gauge fields
\begin{equation}\label{eq:Apmbndcnd}
    A_{+,\bar{w}} \propto \frac{ \nu_+}{\tau} \, ,
    \qquad
    \bar{A}_{+,w} \propto \frac{ \bar{\nu}_+}{\tau} \, ,
    \qquad
    A_{-,\bar{w}} \propto \frac{ \nu_-}{\tau} \, ,
    \qquad
    \bar{A}_{-,w} \propto \frac{ \bar{\nu}_-}{\tau_{}} \, .
\end{equation}
Then the additional holonomy constraints analogous to~\eqref{eq:holcnstr} generate the following set of configurations for the remaining field components
\begin{equation}\label{eq:3DApmshft}
    A_{+,{w}} \propto \frac{ \nu_+}{\tau} + 2m_+ \, ,
    \qquad
    \bar{A}_{+,\bar{w}} \propto \frac{ \bar{\nu}_+}{\tau} + 2\bar{m}_+ \, ,
    \qquad
    A_{-,{w}} \propto \frac{ \nu_-}{\tau} + 2m_- \, ,
    \qquad
    \bar{A}_{-,\bar{w}} \propto \frac{ \bar{\nu}_-}{\tau_{}} + 2\bar{m}_- \, ,
\end{equation}
with $m_\pm, \, \bar{m}_\pm\in \mathbb{Z}$. Plugging these into the equations for the boundary charges, we find the spectral flow equations for two $\sut$ current algebras with levels $k_\pm$
\begin{equation}\label{eq:sfdeflarge}
    \begin{split}
	L_{0} & \rightarrow L_{0}
        + 2\,m_+ \ J^{+,3}_{0} + k_+ \ m_+^2
        + 2\,m_- \ J^{-,3}_{0} + k_- \ m_-^2 \, , \\
	J^{+,3}_{n} & \rightarrow J^{+,3}_{n}
        +  k_+ \ m_+ \ \delta_{n,0} \, , \\
        J^{-,3}_{n} & \rightarrow J^{-,3}_{n}
        +  k_- \ m_- \ \delta_{n,0}\,, \\
        J^{\pm}_n & \rightarrow J^{\pm}_{n \pm m_\pm} \, .
    \end{split}
\end{equation}
This form of spectral flow agrees with what was used in the representation theory analysis of the \emph{large} $\mathcal{N}=4$ algebra in \cite{Petersen:1990a,Petersen:1990b}, which justifies our choice of gauge field boundary conditions, which ultimately led to \eqref{eq:sfdeflarge}.

\subsection{The large \texorpdfstring{$\mathcal{N}=4$}{N4} characters from supergravity}\label{sec:S3S1char}

To find the full partition function of the $j=0$ level of~\eqref{eq:sugraSMS3S1} we proceed as before. We will separately write the contributions of~\eqref{eq:gravqnumlarge1} and \eqref{eq:gravqnumlarge2} for easier comparison to the characters. Using the formulas from \cref{sec:hkapp}, and putting together the partition functions of all fields in~\eqref{eq:gravqnumlarge1} we find (employing the compact notation of \eqref{eq:Z3/2largecmprs})
\begin{equation}\label{eq:lZseed1}
    \begin{split}
        \mathrm{Z}_{(0,0)}^{\text{1-loop}} & = \underbrace{
        \prod_{n=1}^\infty
        \prod_{\tilde{\lambda}_\pm\in\{\pm\}}
        \Big|
        1-z_+^{\tilde{\lambda}_+/2}\,z_-^{\tilde{\lambda}_-/2}\,q^{n-1/2}
        \Big|^2
        }_{4\text{ spin-$\frac{1}{2}$ fermions}} \, 
        \times 
        \underbrace{
        \frac{1}{\prod_{n=1}^\infty
        \Big| 1-q^n \Big|^2}}_{\mathrm{u}(1)\text{ vector}} \, .
    \end{split}
\end{equation}
For the fields of \eqref{eq:gravqnumlarge1} we get
\begin{equation}\label{eq:lZseed2}
    \begin{split}
        \mathrm{Z}_{(0,0)}^{\text{1-loop}} & =
        \underbrace{
        \frac{1}{
        \prod_{n=1}^\infty \prod_{\mu_+\in\{0,\pm\}}
        \Big| 1 - z_+^{\mu_+} \, q^n \Big|^2}
        }_{\text{3 }\sut_+\text{ vectors}} \times
        \underbrace{\frac{1}{
        \prod_{n=1}^\infty \prod_{\mu_-\in\{0,\pm\}}
        \Big| 1 - z_-^{\mu_-} \, q^n \Big|^2}
        }_{\text{3 }\sut_-\text{ vectors}} \\ &
        \times \underbrace{
        \prod_{n=2}^\infty \prod_{\lambda_\pm\in\{\pm\}}
        \Big|
        1-z_+^{\lambda_+/2} \, z_-^{\lambda_-/2} \, q^{n-\frac{1}{2}}
        \Big|^2
        }
        _{\text{4 gravitini}}
        \times \underbrace{\frac{1}{\prod_{n=2}^\infty
        \Big| 1-q^n \Big|^2}}_{\text{graviton}} \, .
    \end{split}
\end{equation}

To this we add the contributions from all other spectral flow sectors $(m_\pm,\bar{m}_\pm)$, which are obtained from~\eqref{eq:lZseed1} and~\eqref{eq:lZseed2} by performing the $\nu_\pm\rightarrow\nu_\pm+2\,m_\pm\,\tau$ shifts, and adding the resulting semi-classical shifts. This results in the full supergravity partition function for the massless fields coming from the $\asx{\mathbf{S}^3\times\mathbf{S}^1}$ compactification
\begin{equation}\label{eq:lZsf}
    \begin{split}
        & \mathrm{Z}_{\text{SUGRA}}^{(1\,\text{loop})} =
        \sum_{m_\pm,\bar{m}_\pm=0}^{\infty} \,
        q^{k_+ m_+^2 +k_+ \bar{m}_+^2 + k_-m_-^2 + k_-\bar{m}_-^2} \,
        z^{k_+m+k_+\bar{m}_++k_-m_-+k_-\bar{m}_-}
        \\ &
        \times
        \frac
        {
        \prod_{n=2}^\infty \prod_{\lambda_\pm\in\{\pm\}}
        \Big|1-z_+^{\lambda_+/2} \, z_-^{\lambda_-/2} \,
        q^{m_++m_--1/2}\Big|^2 \,
        \prod_{n=1}^\infty
        \prod_{\tilde{\lambda}_\pm\in\{\pm\}}
        \Big|1-z_+^{\tilde{\lambda}_+/2} \,
        z_-^{\tilde{\lambda}_-/2} \,
        q^{m_++m_--1/2}\Big|^2 \
        }
        {
        \prod_{n=1}^\infty
        \prod_{\mu_\pm\in\{0,\pm\}}
        \Big| 1 - z_\pm^{\mu_\pm} \, q^{2m_\pm+n} \Big|^2 
        \prod_{n=2}^\infty 
        \Big| 1-q^n \Big|^2 
        \prod_{n=1}^\infty 
        \Big| 1-q^n \Big|^2
        } \, .
    \end{split}
\end{equation}

To compare to the characters of the \emph{large} $\mathcal{N}=4$ algebra, we start by comparing the $m_\pm,\bar{m}_\pm=0$ sectors. As we explained earlier, the algebra ${\mathcal{A}}_{\gamma}$ can be decomposed into two subfactors $\widetilde{\mathcal{A}}_{\gamma}$ and ${\mathcal{A}}_{\text{QU}}$. In~\cref{sec:largealg} we showed that the character of ${\mathcal{A}}_{\gamma}$ also decomposes into characters of the two subalgebras. We will find that the $m_\pm=0$ sector of each of these characters matches onto the holomorphic part of \eqref{eq:lZseed1} and \eqref{eq:lZseed1}, respectively, and that together they give \eqref{eq:lZsf}, after implementing spectral flow. We first copy the $m_\pm=0$ piece of the character of $\widetilde{\mathcal{A}}_{\gamma}$ from~\eqref{eq:lg4chardef}
\begin{equation}\label{eq:lcharseed}
    \begin{split}
        \prod_{n=1}^\infty
        &
        \frac
        {
        (1-z_+^{1/2} \,z_-^{1/2}\,q^{n-1/2}) \,
        (1-z_+^{-1/2}\,z_-^{1/2}\,q^{n-1/2}) \,
        (1-z_+^{1/2}\,z_-^{-1/2}\,q^{n-1/2}) \,
        (1-z_+^{-1/2}\,z_-^{-1/2}\,q^{n-1/2})
        }{
        (1 - z_+ \, q^{n}) \,
        (1 - z_+^{-1} \, q^{n}) \,
        (1 - z_- \, q^{n}) \,
        (1 - z_-^{-1} \, q^{n}) \,
        (1-q^n)^3 \,
        (z_+^{1/2}-z_+^{-1/2}) \,
        (z_-^{1/2}-z_-^{-1/2})} \\ &
        \times
        \left[
        \frac{z_+^{1/2} \, z_-^{1/2}}
        {1+z_+^{1/2}\,z_-^{1/2}\,q^{1/2}} -
        \frac{z_+^{-1/2}\,z_-^{1/2}}
        {1+z_+^{-1/2}\,z_-^{1/2}\,q^{1/2}} -
        \frac{z_+^{1/2}\,z_-^{-1/2}}
        {1+z_+^{1/2}\,z_-^{-1/2}\,q^{1/2}} +
        \frac{z_+^{-1/2}\,z_-^{-1/2}}
        {1+z_+^{-1/2}\,z_-^{-1/2}\,q^{1/2}}
        \right] \, .
    \end{split}
\end{equation}
The factor in the square brackets simplifies to
\begin{equation}\label{eq:lcharalg}
    \frac
    {
    (z_+-z_+^{-1})\,(z_--z_-^{-1})\,(1-q)
    }
    {
    (1+z_+^{}z_-^{}q^{1/2})
    (1+z_+^{-1}z_-^{}q^{1/2})
    (1+z_+^{}z_-^{-1}q^{1/2})
    (1+z_+^{-1}z_-^{-1}q^{1/2})
    } \, ,
\end{equation}
which exactly cancels the required terms in the the first factor of \eqref{eq:lcharseed} to give~\eqref{eq:lZseed2}. The character of ${\mathcal{A}}_{\text{QU}}$ \eqref{eq:lg4charQU} does not have a spectral flow sum, and can directly be matched onto~\eqref{eq:lZseed1}. Therefore, we have a match in the $m_\pm=0$ sector.

The final step is to show that the full character can be obtained from the $m_\pm=0$ piece by spectral flow as defined in~\eqref{eq:sfdeflarge}. For this we use the Theta function form of the full character \eqref{eq:lg4chartheta}, and we copy the $m_\pm=0$ term:
\begin{equation}\label{eq:lcharseedtheta}
    \schr{} (0) \Big|_{m_\pm=0} =
    q^{\frac{5}{8}} \,
    \frac{
    \vartheta_{4}^2 \, (z_+^{1/2}\,z_-^{1/2}, q) \
    \vartheta_{4}^2 \, (z_+^{1/2}\,z_-^{-1/2}, q)
    }{
    \eta(q)^3 \
    \vartheta_1 (z_+, q) \
    \vartheta_1(z_-,q)
    }
    \sum_{\epsilon_{\pm} = \pm 1}
    \epsilon_{+} \, \epsilon_{-} \,\mu(q,z_{+}^{\epsilon_{+}},z_{-}^{\epsilon_{-}})\Big|_{m_\pm=0} \, .
\end{equation}
Performing spectral flow on the character and summing over the sectors amounts to
\begin{equation}\label{eq:lsfonchar}
    \mathfrak{f}(q,z) \rightarrow \sum_{m_\pm\in \mathbb{Z}}\, q^{k_+\,m_+^2+k_-m_-^2}\, z^{k_+\, m_+ + k_-\, m_-} \, \mathfrak{f}(q,z_+ \, q^{2m_+},z_- \, q^{2m_-}) \, .
\end{equation}
We first perform $f(q,z_+,z_-)\rightarrow f(q,z_+\,q^{2m_+},z_-\,q^{2m_-})$ on the second factor in \eqref{eq:lcharseedtheta}. One can check that this reproduces the $m_\pm$-th term in the full character
\begin{equation}\label{eq:lmushft}
    \mu(q,z_+,z_-)\Big|_{m_\pm=0}
    \rightarrow \,
    \mu(q,z_+,z_-)\Big|_{m_\pm} \, ,
\end{equation}
after the $+-$ and $-+$ terms have the sign of $m_+$ and $m_-$ flipped, respectively. Next, we perform the shift on the Theta function factor of \eqref{eq:lcharseedtheta}, and we find that this combination is invariant
\begin{equation}\label{eq:lthetashft}
    \frac{
    \vartheta_{4}^2 \, (z_+^{1/2}\,z_-^{1/2}, \tau) \
    \vartheta_{4}^2 \, (z_+^{1/2}\,z_-^{-1/2}, \tau)
    }{
    \eta(\tau)^3 \
    \vartheta_1 (z_+, \tau) \
    \vartheta_1(z_-,\tau)
    }
    \rightarrow
    \frac{
    \vartheta_{4}^2 \, (z_+^{1/2}\,z_-^{1/2}, \tau) \
    \vartheta_{4}^2 \, (z_+^{1/2}\,z_-^{-1/2}, \tau)
    }{
    \eta(\tau)^3 \
    \vartheta_1 (z_+, \tau) \
    \vartheta_1(z_-,\tau)
    } \, .
\end{equation}
Note that for this to work, we needed the full algebra character, not just that of $\widetilde{\mathcal{A}}_{\gamma}$, which has different powers of the Theta functions. This is because each Theta function, through $\vartheta(q,z \, q^{2m}) = q^{-2m^2} \, z^{-2m} \, \vartheta(q,z)$, produces a non-trivial $(q,z_\pm)$-dependent term after the shift. For all of powers of $(q,z_\pm)$ to cancel, the right number of each Theta functions is required. Had we not included the $\mathcal{A}_{\text{QU}}$ factor, we would not have the squares on the numerator Theta functions, and \eqref{eq:lthetashft} would not hold.

Having matched the partition function and characters at zero spectral flow, and having shown that the remainder of each can be generated through bulk/algebra spectral flow, we find that the full supergravity partition function \eqref{eq:lZsf} and the superconformal character (left and right) \eqref{eq:lg4charfull} match.

\section{Discussion}\label{sec:disc}

In this paper we found the grand canonical partition function of the massless sector of \AdS{3} supergravity obtained by dimensional reduction from 10D spacetime, which itself arises as the low-energy limit of 10D superstring theory. This was done by finding the partition function of each field in the supergravity spectrum. This required us to compute one-loop determinants in \AdS{3} with background gauge fields, which we did using heat kernel methods, extending the known results of pure \AdS{3}. Furthermore, the presence of these gauge fields led to an additional component in the calculation, not present in \AdS{3} theories with less supersymmetry, which is the fact that the path integral boundary conditions of the gauge fields singled out a whole set of configurations. This gave us an answer in the form of a sum over saddles corresponding to different gauge configurations. We showed that the answer we obtained for the partition function matches the product of the left and right characters of the superconformal algebra, the asymptotic algebra of our space. We should mention that we do not claim to have derived the partition function for the first time, as it had already been obtained from the the superconformal characters. Instead we generalized the necessary heat kernel techniques, and used them to derive the partition function working purely in \AdS{3}, and then showed that this agrees with the characters.

\subsection*{Black Holes in \AdS{3}}\label{sec:discBH}

We end by discussing how our work fits into recent calculations of this partition function and related quantities, using methods different from ours, and how these calculations tie to black holes in 3D. A related quantity that can directly be obtained from our result is the partition function in the BTZ black hole background. This relies on the fact that our background, thermal \AdS{3}, and the Euclidean BTZ are the same geometry, but with the time and space circles swapped. As a result, Euclidean BTZ can be obtained from thermal \AdS{3} by relabeling the time and space circles. In terms of the boundary complex structure, this amounts to performing a modular S-transformation $\tau \rightarrow -\frac{1}{\tau}$. This relabeling also causes the fermion boundary conditions on the spatial circle to change. We have been working with periodic fermions in \AdS{3}, which amounts to working in the NS-sector of the superconformal algebra, and with the trace, defined in~\eqref{eq:sm4chardef}, having the $(-1)^F$ fermion number insertion. Performing the modular transformation gives us a trace in the Ramond sector with no $(-1)^F$ insertion. This implies that we can find the BTZ partition function in the Ramond-Ramond sector by performing the S-transform on our result. We summarize this with the following equation
\begin{equation}\label{eq:ZAdStoBTZ}
	\mathrm{Z}_{\text{BTZ in R-R}}(\tau,\nu,\bar{\tau},\bar{\nu}) = \mathrm{Z}_{\text{AdS in NS-NS}} \left(-\frac{1}{\tau},\nu\,\tau,-\frac{1}{\bar{\tau}},\bar{\nu}\,\bar{\tau} \right) \, .
\end{equation}
We also could have obtained the BTZ NS sector trace, which would come from S-transforming the \AdS{3} trace in the NS sector with no $(-1)^F$ insertion. We have chosen to focus on the R-R sector, as it preserves SUSY for BTZ, but all of the discussion that follows equally applies to the NS sector.

Interest in the black hole partition function comes from the fact that it encodes quantum gravitational effects, which start appearing at low temperatures, or near extremality. From the partition function the thermodynamics can be obtained, which is further used in calculations of processes such as evaporation and scattering, all of which exhibit signature quantum effects that arise at low temperature. Recent work has shown how the partition function can be computed for near-extremal black holes using the gravitational path integral in $\asx{K3}$ \cite{Heydeman:2020hhw} and $\asx{\mathbf{S}^3\times\mathbf{S}^1}$ \cite{Heydeman:2025b}, using methods different from those employed in this paper. Near extremality, the path integral receives its dominant contributions from the near horizon region. These contributions come from zero modes associated with the symmetry group of the near-horizon region, which emerges in the near-extremal limit. In \AdS{3} with no supersymmetry, this emergent symmetry group is $\slt$, while for our theories with extended supersymmetry, this emergent symmetry is promoted to $\psut$ for $M=K3$, and to $\text{D}(2,1|\alpha)$ for $M=\mathbf{S}^3\times\mathbf{S}^1$, both of which include as their bosonic subgroups $\slt$ plus additional $\sut$'s. The theory that describes the dynamics of these modes is $\mathcal{N}=4$ super-JT gravity, which reduces to a theory of modes living on the boundary of the near horizon region, the small $\mathcal{N}=4$ Schwarzian theory for $M=K3$, and the large $\mathcal{N}=4$ Schwarzian theory for $M=\mathbf{S}^3\times\mathbf{S}^1$. These theories were exactly solved in~\cite{Heydeman:2020hhw} for the $\psut$ algebra and in~\cite{Heydeman:2025a} for $\text{D}(2,1|\alpha)$, and the partition functions were obtained. Both~\cite{Heydeman:2020hhw,Heydeman:2025b} showed that the BTZ partition functions obtained using these near-extremal methods agree with the full partition function obtained by multiplying the superconformal characters, when the low temperature limit is taken. For more details on how this limit is taken, see~\cite{Ghosh:2020} for \AdS{3} and~\cite{Ferko:2024uxi,Murthy:2025} for the supersymmetric extensions.

As far as black hole physics is concerned, we finally mention that these two isometry groups nicely fit into a classification scheme that was recently studied. The authors of~\cite{Heydeman:2025a} found that under reasonable physical assumptions there are finitely many possible effective theories describing the dynamics of the zero modes in the near horizon of near-extremal black holes in supergravity, and that this classification is closely tied to that of superisometry groups. In addition to the small and large $\mathcal{N}=4$ Schwarzian theories based on $\psut$ and $\text{D}(2,1|\alpha)$, there exist the $\mathcal{N}=3$ Schwarzian theory, the $\mathcal{N}=2$ Schwarzian theory associated with $\mathrm{su}(1,1|1)$, and the $\mathcal{N}=1$ Schwarzian theory. This implies that the largest number of supercharges that can be preserved by the black hole is four (in the left sector, and four in the right), which means that the two examples we studied are the theories with the highest symmetry that can be encountered in black hole backgrounds.

\subsection*{Strings in \AdS{3}}\label{sec:discstring}

The supergravity theories we studied appear as low energy limits of the D1-D5 system in Type II String Theory (with NS-NS fluxes turned on in the compact directions). Strings on \AdS{3}, with possibly additional compact factors, are described by an $\slt$ WZW on the string worldsheet, which is exactly solvable (which can also include extra $\sut$ WZW pieces). The one-loop part of the partition function can be computed for these models, which was originally done in~\cite{Maldacena:2001a,Maldacena:2001b} for \AdS{3}, and was recently performed and reanalyzed for $\asx{T^4}$ in~\cite{Ferko:2024uxi} and $\asx{\mathbf{S}^3\times\mathbf{S}^1}$ in~\cite{Murthy:2025}. The authors found that string theory reproduces the spectrum of chiral primary states
which can be found in supergravity ($j=0$ in \eqref{eq:sugraSMT4} and \eqref{eq:sugraSMS3S1}), and showed that the partition function is consistent with the one obtained in supergravity. They also showed that the generators of the asymptotic spacetime algebra, which correspond to a set of vertex operators in the WZW model, appear in the one-loop determinant, showing explicitly how string theory encodes the spacetime symmetries in the spectrum and partition function. Finally, the \AdS{3} partition function can be S-transformed to a BTZ partition function, which was shown to agree with the one obtained from the superconformal characters. Since the string calculations are performed at finite string length, or specifically at finite $\mathrm{k}^2 = \frac{l_{\text{AdS}}}{l_{\text{string}}}$, this shows that the supergravity result continues to hold at finite $\mathrm{k}$.

\subsection*{Acknowledgements}\label{acknowl}

It is a pleasure to thank our advisor, Prof. Mukund Rangamani, for introducing us to the problem studied in this paper, for the many insightful discussions we had throughout the course of this work, and for reading the draft and providing us with useful feedback. IR and LT were supported by U.S. Department of Energy grant DE-SC0009999 and funds from the University of California. IR would like to thank the Kavli Institute for Theoretical Physics (KITP) for their hospitality during the program, “What is string theory? Weaving perspectives together”, which was supported by the grant NSF PHY-2309135 to KITP.

\newpage

\appendix

\section{The superconformal algebras}\label{sec:algebras}

In this appendix, we summarize the defining relations of the $\mathcal{N} =4$ superconformal algebras, and then present their characters, which we use in the main text.

\subsection{The small \texorpdfstring{$\mathcal{N}=4$}{N=4} superconformal algebra}\label{sec:smallalg}

Let us begin with the \emph{small} $\mathcal{N}=4$ algebra, whose generators are $\{L_n, J^i_n, G_r^a, G_r^{\dot{a}}\}$, with $n \in \mathbb{Z}$, and $i=1,2,3$ and $a,\dot{a}=1,2$ being $\sut$ vector and spinor indices, respectively. In the NS-sector $r\in \mathbb{Z} +\frac{1}{2}$ while in the R-sector $r \in \mathbb{Z}$. The defining commutators are 
\begin{equation}\label{eq:smallN4algcomm}
\begin{split}
\left[ L_m, L_n \right] 
&= 
    (m-n)L_{m+n} 
    + \frac{1}{2} \,k \,m \,(m^2 -1)\, \delta_{m+n,0} \,, \\
\left\lbrace G_r^a, G_s^{\dot{a}} \right\rbrace 
&= 
    2 \,\delta^{a \dot{a}}\, L_{r+s} 
    - 2 \,(r-s) \left(\sigma^i \right)^{a \dot{a}}\, J^i_{r+s} 
    + \frac{1}{2} \,k\, (4\,r^2-1)\, \delta_{r+s,0} \,\delta^{a \dot{a}} \,, \\
\left[ J^i_m, J^j_n \right] 
&= 
    i\, \epsilon^{ijk}\, J^k_{m+n} 
    + \frac{1}{2} \,k \,m \,\delta_{m+n,0}\, \delta^{ij}\,, \\
\left[ L_m, G_r^a \right] 
&= 
    \left( \frac{1}{2}\,m-r \right) G^a_{m+r}\,, 
    \\
\left[ L_m, J^i_n \right] 
&= 
    - n \,J^i_{m+n}\,, \\
\left[ J^i_m, G^a_r \right] 
&= 
    - \frac{1}{2} \left( \sigma^i \right)^{a}{}_b \,  G_{m+r}^b\,.
\end{split}
\end{equation}
The global $\sut$ $R-$symmetry algebra is generated by the zero mode charges $J_0^i$.  The currents $J_n^i$ generate an affine $\sut_k$ current algebra at level $k$, with $k \in \mathbb{Z}_{>0}$ by unitarity. Supersymmetry relates $k$ to the central charge of the Virasoro algebra, $c=6\,k$. The commutators of the subalgebra $\psut$, generated by $\{L_{0,\pm1}, J_0^i, G^a_{\pm\frac{1}{2}}, G^{\dot{a}}_{\pm \frac{1}{2}}\}$, are also contained in~\eqref{eq:smallN4algcomm}.

States of this superalgebra $\ket{h,\ell}$ are labeled by two quantum numbers associated with $L_0$ and $J_0^3$.  We will focus on representations in the NS sector, so $r\in \mathbb{Z} +\frac{1}{2}$. A  highest weight state $\ket{h,\ell}$ is as usual annihilated by the positively moded generators and lies in the spin-$\ell$ highest weight representation of $\sut$. It therefore satisfies 
\begin{equation}\label{eq:smN4annih}
\begin{split}
L_0 \ket{h,\ell} 
&= 
    h \ket{h,\ell}\,, 
\qquad 
    J_0^3 \ket{h,\ell} = j \ket{h,\ell}\,, 
\qquad 
    J^+_0 \ket{h,\ell}  = 0\,, \\
J^i_n \ket{h,\ell} 
&= G^a_r \ket{h,\ell} = G^{\dot{a}}_r \ket{h,\ell} = L_n \ket{h,\ell} = 0, \qquad n,r>0\,.\\
\end{split}
\end{equation}
The rest of the supermultiplet can be obtained by acting with the negatively moded generators. Unitary representations of the algebra in the NS-sector are constrained by having the $\sut$ spin bounded
\begin{equation}\label{eq:sm4unitarity}
    h \geq \ell\,.
\end{equation}
States saturating the bound are the short BPS multiplets.

One feature of the representations is the presence of null vectors, which originate from the presence of the affine $\sut_k$ subalgebra. For instance, a highest weight state $\ket{h,\ell}$ not only satisfies the conditions given in~\eqref{eq:smN4annih}, but $(J_0^-)^{2\ell+1}\ket{h,\ell} =0$ by virtue of $\sut$. Furthermore, one can also check that $G^{\dot{2}}_{1/2} G^1_{1/2}(J_{-1}^+)^{k+1-2\ell} \ket{h,\ell} =0$ and similarly at higher levels. Therefore, one has to remove submodules associated with these null states from the spectrum. An explicit analysis was carried out in~\cite{Eguchi:1987sm,Eguchi:1987wf}, which we will exploit and simply quote the result. 

The representation content can be succinctly encapsulated into a superconformal character. Let 
\begin{equation}\label{eq:sm4chardef} 
\schr{}(h,\ell;\delta) 
=
    \Tr_{_\text{NS}}\left(e^{i\pi\,\delta \,F} \, 
    q^{L_0}\, z^{J_0^3} \right) \,,
\end{equation}
be the character in the generic $h\geq \ell$ representation. The character in question is the NS-sector trace, where for generality we allow for the insertion of twist by $e^{i\pi\,F}$ with $F$ being the fermion number. Hence, the parameter $\delta \in \{0,1\}$ imposes antiperiodic and period boundary conditions for fermions, respectively. Since only short multiplets built atop the highest weight state $\ket{\ell,\ell}$ are present in supergravity, we quote here their character derived in~\cite{Eguchi:1987wf} and refer therein for the general character. We have  
\begin{equation}\label{eq:sm4charorigin}
\begin{split}
    \schr{}(\ell; \delta) 
    &= \,
    q^{\ell} \, 
    \frac
    {\prod_{n=1}
    (1+e^{i\pi\,\delta}\,z^{1/2}\, q^{n-1/2})^2 \,(1+e^{i\pi\,\delta}\, z^{-1/2}\,q^{n-1/2})^2}
    {\prod_{n=1}
    (1-q^n)^2 \,
    (1-z^{+1} \, q^n) \,
    (1-z^{-1}\,
    q^{n-1})} \,
    z^{-1/2}
    \Big(
    \mu(q, z; \delta) - 
    \mu(q, z^{-1}; \delta)
    \Big) \, ,
\end{split}
\end{equation} 
with
\begin{equation}\label{eq:sm4mu}
    \mu(q,z; \delta) 
    = 
    \sum_{m \in \mathbb{Z}} 
    \frac{ z^{\,(k+1)\, m + \ell + 1/2} \, 
    q^{(k+1)\,m^2 + (2\,\ell+1)\,m}}
    {\left( 1 + e^{i\pi\,\delta} \, z^{1/2} \, q^{m + 1/2} \right)^2} \, .
\end{equation}
The first factor captures the oscillator contributions, while the aforementioned null states assemble into the contribution captured by $\mu$. We can also rewrite the character in terms of the Dedekind Eta function and Jacobi Theta functions
\begin{equation}\label{eq:sm4chartheta}
\begin{split}
    \schr{}(\ell; \delta) 
    &= - i  
    q^{\ell + 1/4} \, 
    \frac{\vartheta_{3+\delta}^2 \left( z^{1/2}, q \right)}{\eta(q)^3 \, \vartheta_1 (z, q)} \,
    \Big(
    \mu(q, z; \delta) - 
    \mu(q, z^{-1}; \delta)
    \Big)\,,
\end{split}
\end{equation}
where
\begin{equation}\label{eq:etaf}
    \eta(q) = q^{1/24} \, \prod_{n=1}^{\infty} (1-q^n) \, , 
\end{equation}
and
\begin{equation}\label{eq:thetafs}
    \begin{split}
            \vartheta_1(q,z) & =
            - i \, q^{1/8} \, (z^{1/2}-z^{-1/2})\, 
            \prod_{n=1}^{\infty} \,
            (1-q^n) \,
            (1 - z \, q^{m }) \,
            (1 - z^{-1} \, q^{m}),
            \\
            \vartheta_2(q,z)& =
            q^{1/8} \, (z^{1/2}+z^{-1/2})
            \prod_{n=1}^{\infty} \,
            (1-q^n) \,
            (1 + z \, q^{m}) \,
            (1 + z^{-1} \, q^{m}),
            \\
            \vartheta_3(q,z)& =\prod_{n=1}^{\infty} \,
            (1-q^n) \,
            (1 + z \, q^{m - 1/2}) \,
            (1 + z^{-1} \, q^{m-1/2}),
            \\
            \vartheta_4(q,z) &=\prod_{n=1}^{\infty} \,
            (1-q^n) \,
            (1 - z \, q^{m - 1/2}) \,
            (1 - z^{-1} \, q^{m-1/2}).
    \end{split}
\end{equation}
\subsection{The large \texorpdfstring{$\mathcal{N}=4$}{N=4} superconformal algebra}\label{sec:largealg}

%


The \emph{large} $\mathcal{N} =4$ superconformal algebra, also referred to as $\mathcal{A}_\gamma$, is generated by $\{L_n, J_n^{\pm,i}, G_r^a, U_n, Q_r^a\}$, with $n \in \mathbb{Z}$, $i=1,2,3$, $a,b=0,1,2,3$, $r\in \mathbb{Z} +\frac{1}{2}$ in the NS-sector, and $r \in \mathbb{Z}$ in the R-sector. The defining commutators between $\{L_n, J_n^{\pm,i}, G_r^a\}$ are
\begin{equation}\label{eq:lg4comm1}
    \begin{split}
        \left[ L_m, L_n \right] & = (m-n)L_{m+n} + \frac{c}{12} m (m^2 -1) \delta_{m+n,0}, \\
	   \left\lbrace G_r^a, G_s^b \right\rbrace & = 2 \delta_{ab} L_{r+s} + 4 (r-s) i ( \gamma  \left(\alpha^{+, i}\right)^{ab} J^{+,i}_{r+s} \, + \, (1 - \gamma)  \left(\alpha^{-,i}\right)^{ab} J^{-,i}_{r+s} )
       \\ & + \, \frac{c}{12} (4r^2-1)\delta_{r+s,0} \delta^{ab}, \\
        \left[ J^{\pm,i}_m, J^{\pm,j}_n \right] & = i \epsilon^{ijk} J^{\pm,k}_{m+n} + \frac{1}{2} k^{\pm} m \delta_{m+n,0} \delta^{ij}, \\
	   \left[ J^{\pm,i}_m, G^a_r \right] & = i \left(\alpha^{\pm, i}\right)^{ab} G_{m+r}^b \mp \frac{2 k^{\pm}}{k_+ + k_-} m \left(\alpha^{\pm, i}\right)^{ab} Q^b_{m+r},
    \end{split}
\end{equation}
and the commutators between $\{U_n, Q_r^a\}$ and the rest of the generators are
\begin{equation}\label{eq:lg4comm2}
\begin{split}
        \left\lbrace Q^a_r , Q^b_s \right\rbrace & = \frac{k_+ + k_-}{2} \delta^{ab} \delta_{r+s,0}\,,\\
        \left[ U_m , U_n \right] &= \frac{k_+ + k_-}{2} m \delta_{m+n,0}\, ,\\
        \left\lbrace Q^a_r , G^b_s \right\rbrace &= 2 \alpha_{ab}^{+i} J_{r+s}^{+,i} - 2 \alpha_{ab}^{-i} J_{r+s}^{-,i} + \delta^{ab} U_{r+s}\, ,\\
        \left[ J^{\pm ,i}_{m} , Q^a_r \right] &= i \alpha_{ab}^{\pm i} Q_{m+r}^{b},\\
        \left[ U_m , G_r^{a} \right] &= m Q^a_{m+r}\, ,
\end{split}
\end{equation}
where we used the matrices $\alpha_{ab}^{\pm, i}$ defined as
\begin{equation}\label{eq:lg4alpha}
\left(\alpha^{\pm, i}\right)^{ab} = \frac{1}{2} ( \pm \delta_{ia} \delta_{b0} \mp \delta_{ib} \delta_{a0} + \epsilon_{iab} ).
\end{equation}
The $U_n$ are dimension-1 generators, while the four $Q^a_r$ have dimension equal to $\frac{1}{2}$. The global $\sut_\pm$ $R-$symmetry algebra is generated by the zero mode charges $J_{\pm,0}^i$. The currents $J_n^{\pm,i}$ generate affine $\sut_k$ current algebras at levels $k_\pm$, with $k_\pm \in \mathbb{Z}_{>0}$ by unitarity. Supersymmetry relates $k_\pm$ to the central charge and the $\gamma$ parameter as follows
\begin{equation}\label{eq:lg4cgamma}
c = \frac{6 \, k_+\, k_-}{ k_+ + k_-} \,, \qquad 
\gamma = \frac{k_-}{k_+ + k_-} \,.  
\end{equation}
Furthermore, the commutators of the subalgebra $\text{D}(2,1|\alpha)$, generated by $\{L_0,L_{\pm1},J_0^{\pm,i},G_{\pm\frac{1}{2}}^a\}$, are also contained in~\eqref{eq:lg4comm1}.

Since the sets of geenrators $\{L_n,J^{\pm,i}_n, G_r^a \}$ and $\{U_n, Q^a_r\}$ form two distinct subalgebras $\widetilde{\mathcal{A}}_{\gamma}$ and $\mathcal{A}_\text{QU}$, as was shown by \cite{Gunaydin:1989}, the full $\mathcal{A}_\gamma$ algebra can be decomposed into a direct sum
\begin{equation}\label{eq:lg4decomp}
\mathcal{A}_{\gamma} = \widetilde{\mathcal{A}}_{\gamma} \oplus \mathcal{A}_\text{QU},
\end{equation}
with $\widetilde{\mathcal{A}}_\gamma$ having central charge $c-3$. This decomposition is useful because the representation content of $\mathcal{A}_\text{QU}$ is simply that of a free $\mathfrak{u}(1)$ current and four free spin$-\frac{1}{2}$ fermions. Therefore, the non-trivial part of the representation theory is contained in $\widetilde{\mathcal{A}}$.

States of the superalgebra $\ket{h,\ell_+,\ell_-}$ are labeled by three quantum numbers associated with $L_0$ and $J_{\pm,0}^3$. We will focus on $r\in \mathbb{Z} +\frac{1}{2}$. The highest weight state $\ket{h,\ell_+,\ell_-}$ lies in the spin-$(\ell_+,\ell_-)$ highest weight representation of $\sut_\pm$ and satisfies 
\begin{equation}\label{eq:lg4annih}
\begin{split}
L_0 \ket{h,\ell_+,\ell_-} 
&= 
    h \, \ket{h,\ell_+,\ell_-} \,, 
\quad 
    J_0^{\pm, 3} \ket{h,\ell_+,\ell_-} = \ell_{\pm} \, \ket{h,\ell_+,\ell_-}\,, 
\quad 
    J^{\pm, +}_0 \ket{h,\ell_+,\ell_-}  = 0, \\
J^{\pm, i}_n \ket{h,\ell_+,\ell_-} & = G^a_r \ket{h,\ell_+,\ell_-} = L_n \ket{h,\ell_+,\ell_-} = 0\,, \qquad n,r>0 \,.
    \end{split}
\end{equation}
The rest of the supermultiplet is obtained by acting with the negatively moded generators. Unitary representations of in the NS-sector are constrained by the condition
\begin{equation}\label{eq:lg4unitarity}
    h \geq \frac{1}{k_++k_-}\left[
    k_+\ell_-+k_-\ell_++(\ell_+-\ell_-)^2
    \right] \,,
\end{equation}
with the short BPS multiplets saturating the bound.

In general, the character is in the NS-sector is defined as~\cite{Goddard:1988}
\begin{equation}\label{eq:lg4chardef} 
    \schr{}(h,\ell_+,\ell_-;\delta) 
    =
    \Tr_{_\text{NS}}\left(e^{i\pi\,\delta \,F} \, 
    q^{L_0}\, z_{+}^{J^{+,3}_0} z_{-}^{J^{-,3}_0} \right) \,,
\end{equation}
where we have, for simplicity, turned off the chemical potential for $\mathrm{u(1)}$. The null state structure of the representations and the characters were analyzed in~\cite{Petersen:1990a}. Focusing on $\widetilde{\mathcal{A}}_{\gamma}$, the character for a short representation is given by~\cite{Petersen:1990b} 
\begin{equation}\label{eq:lg4charorigin}
    \begin{split}
	 \tilde{\schr{}}(\ell_+,\ell_-;\delta) & = q^{h} \frac{F^{NS} (q,z_{+},z_{-}) \, B^{+-} (q,z_{+},z_{-})}{\prod_{n=1}^{\infty} (1 - q^n) \left( z_{+}^{1/2} - z_{+}^{-1/2} \right) \left(z_{-}^{1/2} - z_{-}^{-1/2}\right)} \\
	& \times \sum_{\epsilon_{\pm} = \pm 1} \epsilon_{+} \epsilon_{-} \mu(q,z_{+}^{\epsilon_{+}},z_{-}^{\epsilon_{-}}),
    \end{split}
\end{equation}
where $h$ satisfies the bound~\eqref{eq:lg4unitarity}. The term
\begin{equation}\label{eq:lg4F}
    \begin{split}
	F^{NS}(q,z_{+},z_{-}) = \prod_{n=1}^{\infty} \prod_{\epsilon_{\pm} \in {\pm}} (1 + e^{i \pi \delta} z_{+}^{\epsilon_{+}/2} z_{-}^{\epsilon_{-}/2} q^{n-1/2})\, ,
    \end{split}
\end{equation}
accounts for contributions from fermionic oscillators, whereas
\begin{equation}\label{eq:lg4B}
    \begin{split}
	B^{+-}(q,z_{+},z_{-}) = \prod_{n=1}^{\infty} \frac{1}{(1 - q^n)^2 (1 - z_{+} q^n) (1 - z_{+}^{-1} q^n) (1 - z_{-} q^n) (1 - z_{-}^{-1} q^n)} \, ,
    \end{split}
\end{equation}
encodes the contributions from bosonic current generators. In addition, the Virasoro oscillators are captured by the factor $\prod_{n=1}^{\infty} (1 - q^n)^{-1}$. Finally, the sum
\begin{equation}\label{eq:lg4mu}
    \begin{split}
	\mu(q,z_{+},z_{-}) = \sum_{m_\pm \in \mathbb{Z}} q^{m_+^2 k_{+} + m_-^2 k_{-} + m_+ (2 l_{+} + 1) + m_- (2 l_{-} + 1)}
    \frac{
    z_{+}^{m_+ k_{+} + l_{+} + 1/2} z_{-}^{m_- k_{-} + l_{-} + 1/2}}{1 +e^{i \pi \delta}\, z_{+}^{1/2} z_{-}^{1/2} q^{m_++m_-+1/2}
    } \, ,
    \end{split}
\end{equation}
captures the contributions from the null states. This character can also be written in terms of Theta and Eta functions as
\begin{equation}\label{eq:lg4chartheta}
	 \tilde{\schr{}}(\ell_+,\ell_-;\delta) = 
     -q^{h + 1/2}
     \frac{
     \vartheta_{3+\delta} \left(q,z_+^{1/2}\,z_-^{1/2}\right) \,
     \vartheta_{3+\delta} \left(q,z_+^{1/2}\,z_-^{-1/2}\right)
     }{
     \eta(\tau)^3 \,
     \vartheta_1 (q,z_+) \,
     \vartheta_1(q,z_-)
     }
     \sum_{\epsilon_{\pm} = \pm}
     \epsilon_{+} \, \epsilon_{-} \,
     \mu(q,z_{+}^{\epsilon_{+}},z_{-}^{\epsilon_{-}})\, .
\end{equation}
On the other hand, the character for a unitary representation of $\mathcal{A}_{QU}$ is given by~\cite{Petersen:1990a}
\begin{equation}\label{eq:lg4charQU}
    \begin{split}
	 \schr{QU}(\ell_+,\ell_-;\delta) = q^{-1/8}\frac{F^{NS} (q,z_{+},z_{-})}{\prod_{n=1}^{\infty} (1 - q^n)}\,,
    \end{split}
\end{equation}
where the numerators accounts for the four fermions, and the denominator for the one bosonic current. For comparison to the references, note that we have dropped the overall central charge dependence $q^{-\frac{c}{12}}$ in the definition of the character. Putting \eqref{eq:lg4chartheta} and \eqref{eq:lg4charQU} together, we find the following equation for the character of the \emph{large} $\mathcal{N}=4$ algebra
\begin{equation}\label{eq:lg4charfull}
\schr{}(\ell_+,\ell_-;\delta) = -q^{h + 1/2} \frac{
     \vartheta_{3+\delta} \left(q,z_+^{1/2}\,z_-^{1/2}\right) \,
     \vartheta_{3+\delta} \left(q,z_+^{1/2}\,z_-^{-1/2}\right)
     }{
     \eta(\tau)^6 \,
     \vartheta_1 (q,z_+) \,
     \vartheta_1(q,z_-)
     }
    \sum_{\epsilon_{\pm} = \pm 1}
    \epsilon_{+} \, \epsilon_{-} \,\mu(q,z_{+}^{\epsilon_{+}},z_{-}^{\epsilon_{-}})\, .
\end{equation}

\section{One-loop partition function on thermal \texorpdfstring{$\asth$}{\asth}}\label{sec:hkapp}

As outlined in~\cref{sec:hkgeneral}, the 6D SUGRA partition function follows from computing the contributions to the one-loop determinant of arbitrary spin$-s$ particles on thermal $\asth$. Performing the $t-$integral in~\eqref{eq:hk6D1hel} yields the following expression for the one-loop determinant
\begin{equation} \label{eq:one-loop-kernel-bare}
    \begin{split}
        &- \frac{1}{2} \ln  \det \left(  - \Delta_{(s)} + M^2  \right) = \\
        & \sum_{m=1}^\infty (-1)^{m (\delta+1)} \frac{ q^{s \, m } \, \chi_{\ell}(z^m) \chi_{\bar{\ell}} ( \bar{z}^m) \, +  \bar{q}^{s \,m} \, \chi_{\bar{\ell}}(z^m) \chi_{\ell} (\bar{z}^m)}{m |1-q^m|^2} \, |q|^{m \, \left( \sqrt{s+1+M^2} - (s-1) \right)} .
    \end{split}
\end{equation}
Naively, we can compute the one-loop partition function by 
\begin{equation} \label{eq:one-loop-determinant-massive}
Z_{1-\text{loop}} = \text{det}^{-1/2}(- \Delta_{(s)}+M^2).
\end{equation}
This equation can be directly applied to scalars, spinor fields and massive higher-spin fields, but not to massless higher-spin fields, the main problem being that the gauge-fixing procedure introduces ghost contributions to the one-loop partition function.

It is well known that the mass of a particle living in \AdS{3} acquires a spin-dependent contribution from the curvature. Therefore, $M$ appearing in the above expression is the total mass, which can be decomposed into the physical mass $m_p$ and the curvature contribution $m_s$:
\begin{equation}
    M^2 = m_p^2 + m_s^2.
\end{equation}
In particular, $m_s$ can be non-zero even in the massless case.

To start, let us consider the case of massive particles, i.e. $m_p \neq 0$. The physical mass is related to the conformal dimension $h$ as~\cite{Aharony:2000, Datta:2011}: 
\begin{equation} \label{eq:mass_spin}
    \begin{split}
        \text{scalar fields}: \quad & h=\frac{1}{2} \left( 1+ \sqrt{1 + m_p^2} \right), \\
        \text{spinor fields}: \quad & h= \frac{1}{2} \left( m_p + \frac{3}{2} \right), \\
        \text{vector fields}: \quad & h=1+ \frac{m_p}{2}, \\
        \text{gravitinos}: \quad & h=\frac{1}{2} \left( m_p + \frac{5}{2} \right), \\
        \text{gravitons}: \quad & h= \frac{1}{2} \left(3 + \sqrt{1+m_p^2} \right).
    \end{split}
\end{equation}
whereas the spin-dependent contribution looks~\cite{Gaberdiel:2010, Creutzig:2012}
\begin{equation}
    m_s^2 =
    \begin{cases}
        s(s-3) & \qquad s=1,\,2, \\
        - \left(s+1 \right) & \qquad  s=\frac{1}{2}, \,\frac{3}{2}.
    \end{cases}
\end{equation}
Then~\eqref{eq:one-loop-kernel-bare} can be rewritten as
\begin{equation}
    \begin{split}
        & - \frac{1}{2} \ln  \det \left(  - \Delta_{(s)} + M^2  \right) = \\
        & \sum_{m=1}^\infty (-1)^{m (\delta+1)} \frac{ q^{\,s m } \, \chi_{\ell}\left( z^m \right) \chi_{\bar{\ell}} \left( \bar{z}^m \right) \, + \bar{q}^{\, s m} \, \chi_{\bar{\ell}} \left( z^m \right) \chi_{\ell} \left( \bar{z}^m \right)}{m |1 - q^m|^2} |q|^{\,2(h-s)} .
    \end{split}
\end{equation}
A straightforward application of~\eqref{eq:one-loop-determinant-massive} yields
\begin{equation}\label{eq:massive_one_loop}
    \begin{split}
        Z_\text{{1-loop}} = \Bigg| \prod_{n, \bar{n}, p, \bar{p}=0}^\infty & \frac{\left( 1+e^{i \pi \delta \,} z^{\, \ell + 1+ p} \bar{z}^{ \, \bar{p} -\bar{\ell} } q^{\, h + n} \bar{q}^{\, h-s+\bar{n}} \right) }{\left( 1+e^{i \pi \delta \,} z^{\, \ell + 1+ p} \bar{z}^{\, \bar{\ell} +1 + \bar{p}} q^{\, h+n} \bar{q}^{\, h-s+\bar{n}} \right)} \\
        & \frac{\left( 1+e^{i \pi \delta \,}  z^{ \, p -\ell} \bar{z}^{\, \bar{\ell} +1 +\bar{p}} q^{\, h+n} \bar{q}^{\, h-s+\bar{n}} \right)}{\left( 1+e^{i \pi \delta \,} z^{\, p-\ell} \bar{z}^{\, \bar{p}- \bar{\ell}} q^{\, h+n} \bar{q}^{\, h-s+\bar{n}} \right)} \Bigg|^{2(-1)^F} .
    \end{split}
\end{equation}
where $\delta = 1$ for bosonic particles and $F$ is the fermion number.

In the case of massless particles the one-loop partition function is affected by the ghost contributions arising from gauge symmetry. We have~\cite{Gaberdiel:2010, Creutzig:2012}
\begin{equation}
    Z_{1-\text{loop}} = \left[ \frac{\text{det}^{\frac{1}{2}} \left(-\Delta_{(s-1)}+s(s-1) \right)^{TT}}{\text{det}^{\frac{1}{2}} \left( -\Delta_{(s)}+s(s-3) \right)^{TT}}\right]^{(-1)^F}, \qquad s \geq 1 
\end{equation}
where, as indicated, the Laplacian acts only on the traceless and transverse degrees of freedom of the spin-$s$ and spin-$(s-1)$ fields. Then
\begin{equation} \label{eq:massless_one_loop}
    \begin{split}
    Z_{1-\text{loop}} = \Bigg| \prod_{n, p, \bar{p}=0}^\infty   \frac{\left( 1+e^{\,i \pi \delta \,} z^{\, \ell + 1+ p} \bar{z}^{ \, \bar{p} -\bar{\ell}} q^{\, n+s} \right) \left( 1+e^{\,i \pi \delta \,}  z^{ \, p -\ell} \bar{z}^{\, \bar{\ell} +1 +\bar{p}} q^{\, n+s} \right)}{\left( 1+e^{\,i \pi \delta \,} z^{\, \ell + 1+ p} \bar{z}^{\, \bar{\ell} +1 + \bar{p}} q^{\, n+s}\right) \left( 1+e^{\,i \pi \delta \,} z^{\, p-\ell} \bar{z}^{\, \bar{p}- \bar{\ell}} q^{\, n+s} \right)} \Bigg|^{2(-1)^F} 
    \end{split}
\end{equation}
In particular we get
\begin{itemize}
    \item $s=2$:
        \begin{equation}
             Z_{1-\text{loop}} = \prod_{n=2}^\infty \frac{1}{|1-q^n|^2}.
        \end{equation}
    \item $s=\frac{3}{2}$:
    \begin{equation}
        Z_{1-\text{loop}} = \prod_{n=2}^\infty \Bigg|  \left( 1 +e^{i \pi \delta}\, z^{-\frac{1}{2}} q^{n - \frac{1}{2}} \right) \left( 1 +e^{i \pi \delta}\, z^{\frac{1}{2}} q^{n - \frac{1}{2}} \right) \Bigg|^2.
    \end{equation}
    \item $s=1$:
    \begin{equation}\label{eq:appZvect}
        Z_{1-\text{loop}} = \prod_{n=1}^\infty \Bigg| \frac{1}{(1-z^{-1} q^n) (1- q^n) (1-z q^n)} \Bigg|^2.
    \end{equation}
\end{itemize}

In order to find the full one-loop partition function for supergravity theory on $\asth$ we glue together~\eqref{eq:massive_one_loop} and ~\eqref{eq:massless_one_loop} as outlined in the main text.

\newpage
\bibliographystyle{JHEP}
\bibliography{N4superchar}

\end{document}